\pdfoutput=1
\documentclass[letterpaper,twocolumn,10pt]{article}
\usepackage{usenix,epsfig,endnotes}
\usepackage{enumitem}
\usepackage{graphicx}
\usepackage{epstopdf}
\usepackage{amssymb}
\usepackage{amsmath,amsthm}
\usepackage{comment}
\usepackage{longtable}
\usepackage{multirow}
\usepackage{booktabs}
\usepackage{pslatex}
\usepackage{tabularx}
\everypar{\looseness=-1}
\usepackage[T1]{fontenc}
\usepackage[compact]{titlesec}

\usepackage{subcaption}
\usepackage{textcomp}
\usepackage{bbding}
\usepackage{titling}
\DeclareRobustCommand*\circled[1]{\tikz[baseline=(char.base)]{ \node[shape=circle,draw,color=white,fill=black,inner sep=0.5pt] (char){#1};}}

\def\ie{{i.e.},~}
\def\eg{{e.g.},~}

\newenvironment{nstabbing}
  {\setlength{\topsep}{0pt}%
   \setlength{\partopsep}{0pt}%
   \tabbing}
{\endtabbing}

\usepackage[noadjust]{cite}
\usepackage{tikz}
\usepackage{pifont}

\usepackage[flushleft]{threeparttable}
\usepackage[linesnumbered,ruled]{algorithm2e} 

\usepackage{microtype}
\usepackage[font={small}]{caption}
\usepackage{setspace}
\usepackage[scaled=.8]{beramono}

\usepackage{listings}
\usepackage{color}
\DeclareCaptionFont{black}{\color{black}}
\usepackage{xspace}
\newcommand{\Soteria}{{\textsc{\footnotesize{Soteria}}}\xspace}
\newcommand{\MalIoT}{{\textsc{\footnotesize{MalIoT}}}\xspace}
\usepackage{authblk}

\newcommand{\setanonymous}[1]{
    \newboolean{anonflag}
    \setboolean{anonflag}{#1}
}
\newcommand{\anonymous}[1]{
    \ifthenelse{\boolean{anonflag}}{}{#1}
}
\setanonymous{false}

\newcommand{\showextended}{1} 
\newcommand{\extended}[1]{\ifthenelse{\equal{\showextended}{1}}{#1}{}}

\usepackage{setspace}
\usepackage{etoolbox}
\usepackage{tikzpagenodes}

\begin{document}

\title{\textsc{Soteria}: Automated IoT Safety and Security Analysis}
\anonymous{
\setcounter{Maxaffil}{1}
\author{Z. Berkay Celik}
\author{Patrick McDaniel}
\author{Gang Tan}
\affil{Department of Computer Science and Engineering\\The Pennsylvania State University\authorcr
    {\texttt{\{zbc102,mcdaniel,gtan\}}@cse.psu.edu}}
}

\ifthenelse{\boolean{anonflag}}{
    \author{\vspace{-1.2em}
    Anonymous}
}{}

\setlength{\droptitle}{-20pt}
\date{}

\maketitle
\begin{tikzpicture}[remember picture,overlay]
    \node[align=center] at ([yshift=0.5em]current page text area.north) {Extended version of the Soteria: Automated IoT Safety and Security Analysis \\ (Accepted to the  USENIX Annual Technical Conference (USENIX ATC), 2018).};
  \end{tikzpicture}%
 
\thispagestyle{plain}
\pagestyle{plain}

\vspace{-25pt}

\begin{abstract}
Broadly defined as the Internet of Things (IoT), the growth of commodity devices that integrate physical processes with digital systems have changed the way we live, play and work. Yet existing IoT platforms cannot evaluate whether an IoT app or environment is safe, secure, and operates correctly. In this paper, we present \Soteria, a static analysis system for validating whether an IoT app or IoT environment (collection of apps working in concert) adheres to identified safety, security, and functional properties. \Soteria operates in three phases; (a) translation of platform-specific IoT source code into an intermediate representation (IR), (b) extracting a state model from the IR, (c) applying model checking to verify desired properties. We evaluate \Soteria on 65 SmartThings market apps through 35 properties and find nine (14\%) individual apps violate ten (29\%) properties. Further, our study of combined app environments uncovered eleven property violations not exhibited in the isolated apps. Lastly, we demonstrate \Soteria on \MalIoT, a novel open-source test suite containing 17 apps with 20 unique violations.
\end{abstract}

\section{Introduction} 
The introduction of IoT devices into public and private spaces has changed the way we live. For example, home automation apps supporting smart devices of thermostats, locks, switches, surveillance systems, and Internet-connected appliances change the way we interact with our living spaces.  While these systems have been widely embraced, IoT has also raised concerns about the security and safety of digitally augmented lives~\cite{orc16, ronen2017iot, fernandes2016security, ho2016smart, jab15}. IoT apps have access to functions that may put the user or environment at risk, \eg unlock doors when not at home or create unsafe or damaging conditions by turning off the heat in winter. There has been an increasing amount of recent research exploring IoT security and more broadly environmental safety. 

One of the oft-discussed criticisms of IoT is that the software and hardware frameworks do not possess the capability to determine if an IoT device or environment is implemented in a way that is safe, secure, and operates correctly. The SmartThings~\cite{samsung}, OpenHab~\cite{openhab}, Apple's Homekit~\cite{apple} provide guidelines and policies for regulating security~\cite{Official, OpenHabGuideline, appleSecurity}, and related markets provide a degree of internal (hand) vetting of the apps prior to distribution~\cite{smartThings-review, AppleHomekitReview}.  Recent technical community efforts have explored vulnerability analysis within targeted IoT domains~\cite{oluwafemi2013experimental, ho2016smart}, while others focused on sensitive data leaks and correctness of IoT apps using a range of analyses~\cite{fernandes2016flowfence, saint-taint-analysis, jia2017contexiot, tian2017smartauth}. However, tools and algorithms for evaluating general safety and security properties within IoT apps and environments are at this time largely absent. 

In this paper, we present \Soteria\footnote{Soteria is the goddess in Greek mythology preserving from harm.}, a static analysis system for validating whether an IoT app or IoT environment (collection of apps working in concert) adheres to identified safety, security, and functional properties. We exploit existing IoT platforms' sensor-computation-actuator program structures to translate source code of an IoT app into an intermediate representation (IR). Here, the \Soteria IR models the app's lifecycle--including app entry points, event handler methods, and call graphs. From this, \Soteria uses the IR to perform efficient static analysis extracting a state model of the app; the state model includes its states and transitions. A set of IoT properties is systematically developed, and model checking is used to check that the app (or collection of apps) conforms to those properties. In this work, we make the following contributions:

\begin{itemize}[noitemsep,nolistsep]
\item We introduce \Soteria, a system designed for model checking of IoT apps. \Soteria automatically extracts a state model from a SmartThings IoT app and applies model checking to find property violations.  

\item We used \Soteria on 65 different IoT apps (35 apps from the official SmartThings repository and 30 community-contributed third-party apps from the official SmartThings forum) and reveal how safety and security properties are violated.

\item We develop an IoT-specific test corpus \MalIoT, an open-source repository of 17 flawed apps that containing an array of safety and security violations.
\end{itemize}

\begin{figure}[t!]
\begin{center}
\includegraphics[width=1\columnwidth]{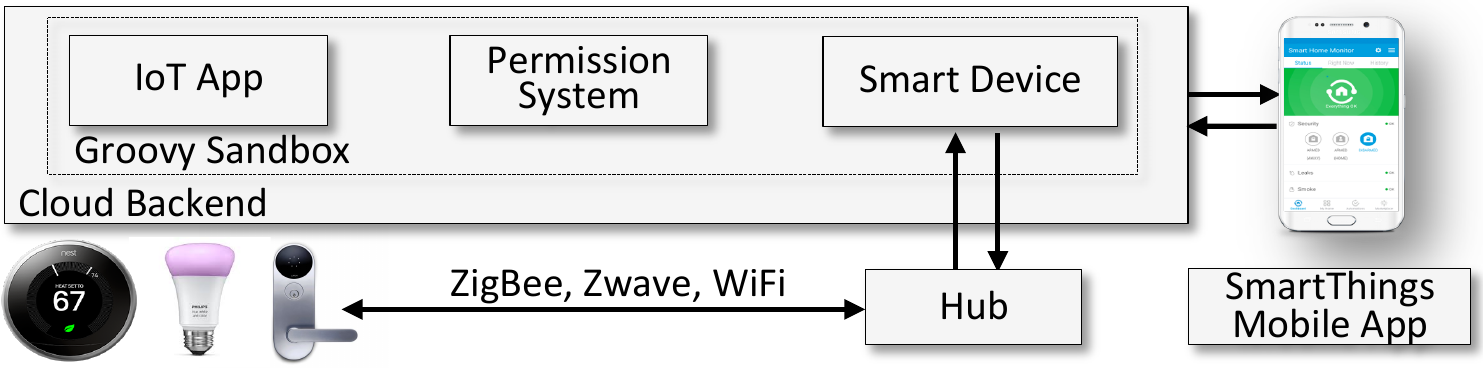}
\caption{The architecture of SmartThings IoT platform.}
\label{fig:smartthing-architecture}
\end{center}
\end{figure}

\section{Background}
\noindent \textbf{IoT platforms} provide a software stack used to develop applications that monitor and control IoT devices.\footnote{While the \Soteria approach is largely agnostic to the specific IoT software platform, we focus on Samsung's SmartThings Platform~\cite{samsung}.} For example, Fig.~\ref{fig:smartthing-architecture} shows the three components of the Samsung's SmartThings Platform: a hub, apps, and the cloud backend~\cite{smartThings-review}. The hub controls the communication between connected devices, cloud back-end, and mobile apps. Apps are developed in the  Groovy language (a dynamic, object-oriented language) and executed in a Kohsuke sandboxed environment. The cloud backend creates software proxies called SmartDevices that act as a conduit for physical devices, as well as run the apps.

The permission system in SmartThings allows a developer to specify devices and user inputs required for an app at install time. Devices in SmartThings have capabilities (\ie permissions) that are composed of \emph{actions} and \emph{events}. Actions represent how to control or actuate device states and events are triggered when device states change. SmartThings apps control one or more devices.  Apps subscribe to device events or other pre-defined events such as the icon-clicking event, and an event handler is invoked to handle it, which may lead to further events and actions. 

Users can install SmartThings apps either from the market or proprietary system via SmartThings Mobile.  In the former, publishing an app in the official market requires the developer to submit the source code of the app for review. Official apps appear in the market after the completion of a lengthy review process~\cite{smartThings-review}. In the latter, organizations can develop an app and make it accessible using the Web IDE. These self-published apps do not receive any official review process and are often shared in the SmartThings official community forum~\cite{Community}. 

\extended{
\vspace{2pt}\noindent \textbf{Formal Program Verification} is used as a vehicle to analyze the correctness of software in safety-critical systems concerning a formal property~\cite{jhala2009software}. The two most popular methods for formal verification are automated theorem proving and model checking. In this paper, we apply verification based on model checking. In model checking, programs are represented as a state-machine model and the execution of the software is checked against properties. A temporal formula is used to define a property to be verified, and it is checked against the state-machine model using a model checker.

The properties can be written in temporal logic formulas such as Linear Temporal Logic (LTL) or Computational Tree Logic (CTL)~\cite{clarke1981design}, which combines path quantifiers with linear-time operators, making it amenable to state-based model checking. Model checkers automate verification after a model is constructed and properties are specified. In practice, a model checker is chosen based on a particular set of verification requirements~\cite{zhang2010classification}.
}

\begin{figure}[t!]
\begin{center}
\includegraphics[width=1\columnwidth]{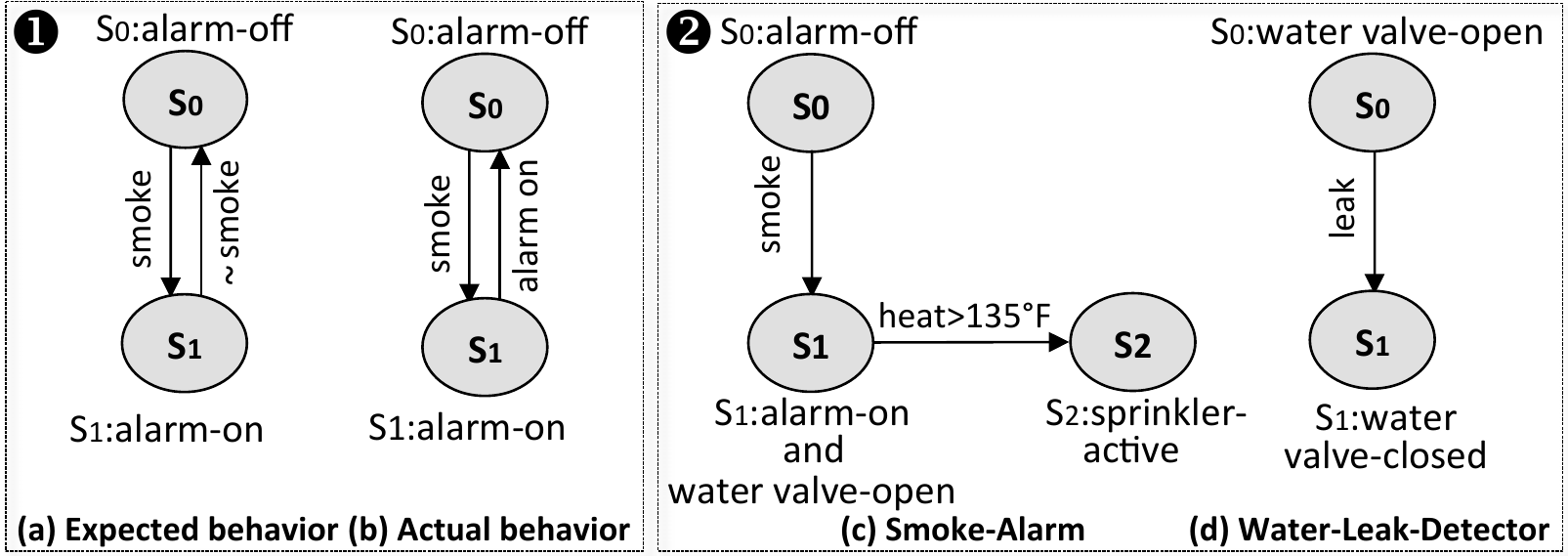}
\caption{\circled{1} shows the state models of the expected and actual behavior of the \texttt{Smoke-Alarm} app. The app fails because of a bug which halts the alarm when smoke is present. \circled{2} shows the state models of the \texttt{Smoke-Alarm} and \texttt{Water-Leak-Detector} apps violating a property when they installed together. The environment fails when the apps interact--the \texttt{Water-Leak-Detector} app shuts off water valve when a fire is detected.}
\label{fig:motivating-example}
\end{center}
\end{figure}

\section{Motivation and Assumptions}
\label{sec:problem-attack}
\noindent\textbf{Example IoT Applications.} We introduce three running examples used throughout for exposition and illustration \extended{(We present the source code of the apps in Appendix~\ref{appendix:example-app})}:

\vspace{2pt}\noindent\textbf{The \texttt{Smoke-Alarm} app} contains a smoke-detection alarm, a water valve (basement), and a light switch (living room). The app sounds the smoke alarm and turns on the light when smoke is detected; when smoke is detected and a heat level is reached, the app opens the water valve to activate fire sprinklers; finally, it turns off the alarm and closes water valve when smoke is clear. Also, it turns on the light switch when the smoke-detector battery is low.

\vspace{2pt}\noindent\textbf{The \texttt{Water-Leak-Detector} app} detects a water leak with a moisture sensor and shuts off the main water supply valve in order to prevent any further water damage. 

\vspace{2pt}\noindent\textbf{The \texttt{Thermostat-Energy-Control} app} locks the front door and sets the heating thermostat temperature to a pre-defined value when the user-presence mode is changed (\eg from the user-away mode to the user-home mode or vice versa). When the energy usage is above a pre-defined consumption threshold, it turns off the thermostat switch.

\vspace{2pt}\noindent\textbf{\Soteria illustrated.} Here we informally illustrate \Soteria analysis through a single and multi-app example.

Consider the \texttt{Smoke-Alarm} app. We first model the app's source code as a transition system. Fig.~\ref{fig:motivating-example}(1a) presents the expected behavior of the smoke alarm; the alarm sounds when smoke is detected and not otherwise. The state model starts from an initial state $\mathtt{S_0}$ and transits to state $\mathtt{S_1}$ when smoke is detected. The state transitions are controlled by the output of the smoke sensor: ``smoke-detected'' (smoke) and ``not detected'' (\texttildelow smoke). Fig.~\ref{fig:motivating-example}(1b) is the actual behavior extracted from the open-source implementation of a smoke alarm (that has a bug). We use \Soteria to validate the above safety property--i.e., ``does the alarm always sound when there is smoke?'' To perform this analysis \Soteria encodes the safety property in temporal logic and verifies it on the model with a symbolic model checker.  Naturally, the analysis showed a violation; the actual behavior of the app stops the sound moments after the alarm sounds (the state transition from $\mathtt{S_1}$ to $\mathtt{S_0}$). In this case, users may not hear the short or intermittent alarm with potentially disastrous consequences. 

Now consider the situation when both \texttt{Smoke-Alarm} and \texttt{Water-Leak-Detector} apps are co-located in an environment. Fig.~\ref{fig:motivating-example}(2c) and~\ref{fig:motivating-example}(2d)  presents expected behavior of the \texttt{Smoke-Alarm} and \texttt{Water-Leak-Detector} apps, respectively.  Here, we use \Soteria to validate the property ``does the sprinkler system activate when there is a fire?''. The model checker revealed that there was a safety violation: the \texttt{Water-Leak-Detector} app shuts off the water valve and stops fire sprinklers when it detects water release from sprinklers. In this case, the joint behavior of the otherwise-safe apps leaves users are at risk from fire. 

\noindent\textbf{Assumptions and Threat Model.} We assume violations can be caused by design flaws or malicious intent.  In the latter, the adversary may insert malicious code resulting in insecure or unsafe states, e.g., as seen in attacks on smart light bulbs~\cite{ronen2017iot} and home security systems~\cite{ronen2016extended}. We do not evaluate adversaries' ability to thwart security measures (e.g., crypto, forged inputs) or explore user privacy, but defer those investigations to future work.

\section{\textsc{Soteria}}
\begin{figure}[t!]
\begin{center}
\includegraphics[width=1\columnwidth]{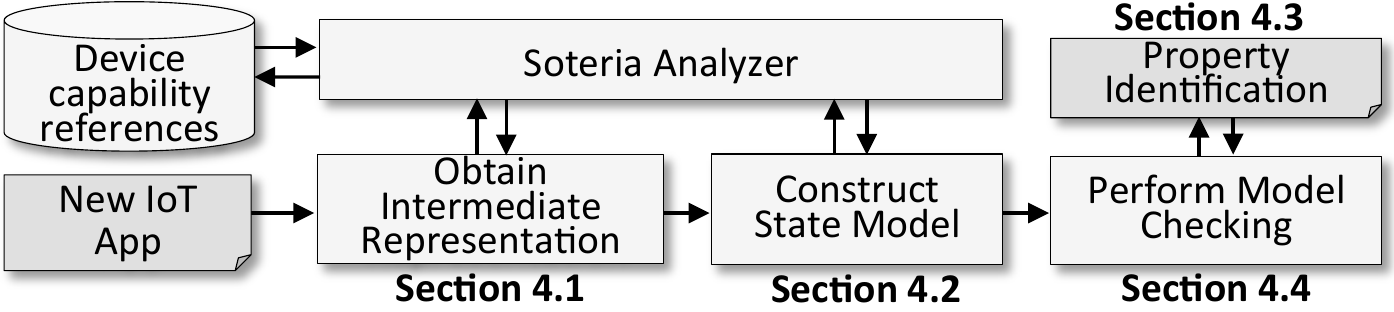}
\caption{Overview of \Soteria{} architecture.}
\label{fig:methodology}
\end{center}
\end{figure}

Fig.~\ref{fig:methodology} provides an overview of the four stages of the \Soteria system analysis.  \Soteria first extracts an intermediate representation (IR) from the source code of an IoT app (Sec.~\ref{sec:IR}). The IR is used to model the lifecycle of an app including entry points, event handler methods, and call graphs. Second, \Soteria uses the IR to extract a state model of the app; the state model includes its states and transitions (Sec.~\ref{sec:fsm-extraction}). Lastly, a set of IoT properties is developed (Sec.~\ref{sec:property-identification}), and model checking is used to check that the app conforms to those properties when running independently or interacting with other apps (Sec.~\ref{sec:property-validation}). 

\subsection{From Source Code to IR}
\label{sec:IR}
\begin{figure}[t!]
\begin{center}
\includegraphics[width=1\columnwidth]{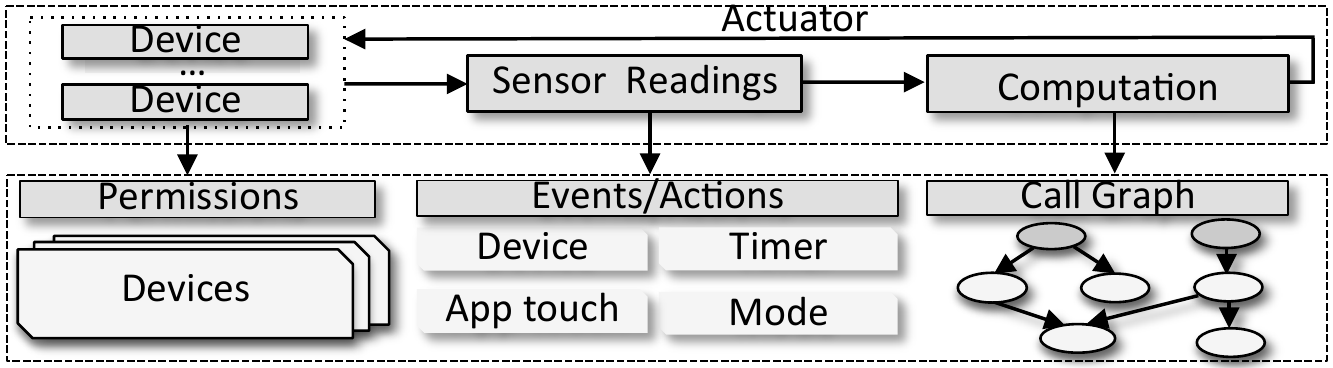}
\caption{Components of the intermediate representation (IR).}
\label{fig:components}
\end{center}
\end{figure}

The first step toward modeling an IoT app is to extract an IR from the app's source code. \Soteria builds the IR from a framework-agnostic component model, which is comprised of the building blocks of IoT apps, shown in Fig.~\ref{fig:components}. A broad investigation of existing IoT environments showed that the programming environments could be generalized into three component types: (1) \textit{Permissions} grant capabilities to devices used in an app; (2) \textit{Events/Actions} reflect the association between events and actions: when an event is triggered, an associated action is performed; and (3) \textit{Call graphs} represent the relationship between entry points and functions in an app. 
The IR has several benefits. First, it allows us to precisely model the app lifecycle as described above.  Second, it is used to abstract away parts of the code that are not relevant to property analysis, e.g., \texttt{definition} blocks that specify app meta-data and \texttt{logger} logging code. Third, it allows efficiently extract the states and state transitions from the implementation (see below). Presented in Fig.~\ref{fig:IRcode}, we use the \texttt{Smoke-Alarm} app to illustrate the use of the IR. 

\vspace{2pt}\noindent\textbf{Permissions.} When a SmartThings app gets installed or updated, the permissions for devices and user inputs are displayed to the user (and explicitly accepted). The permissions are read-only, and app logic is implemented using the permissions. \Soteria visits permissions of an app to extract its devices and user inputs. Turning to the IR in Fig.~\ref{fig:IRcode}, the permission block (lines 1--7) defines: (1) the devices; a smoke detector, a switch, an alarm, a valve, and a battery in the smoke detector; and (2) user input: ``thrshld'' is used to determine whether the battery level of the smoke detector is low. For each permission, the IR declares a triple following keyword ``input''. For a device, the triple associates an identifier for the device, called the \emph{device handle}, to its platform-specific device name in order to determine the interface that the device may access. For instance, an app may associate identifier \texttt{the\_switch} with a switch device, which is in either the ``off'' or the ``on'' state. For a user input, the triple in the IR contains the variable name storing the user input, its type, and a tag showing the kind of input such as the user-defined input. In this way, we obtain a complete list of devices and user inputs that an app might access. 

\vspace{2pt}\noindent\textbf{Events/Actions.} Similar to mobile applications, an IoT app does not have a main method due to its event-driven nature. Apps implicitly define entry points by subscribing events. The event/actions block in an IR is built by analyzing how an app subscribes to events. Each line in the block includes three pieces of information: a device handle, a device event to be subscribed, and an event handler method to be invoked when that event occurs (lines 9--10). Event handler methods are commonly used to take device actions. Therefore, an app may define multiple entry points by subscribing multiple events of a device or devices. Turning to our example, the event of ``smoke-detected'' state change is associated with an event handler method named \texttt{h1()} and the event of ``battery'' level state change with \texttt{h2()}. We also found that events are not limited to device events; we call these \emph{abstract events}: (1) \emph{Timer events}; event-handlers are scheduled to take actions within a particular time or at pre-defined times (\eg an event-handler is invoked to take actions after a given number of minutes has elapsed or at specific times such as sunset); (2) \emph{App touch events}; for example, some action can be performed when the user clicks on a button in an app; (3) what actions get generated may also depend on \emph{mode events}, which are behavior filters automating device actions. For instance, an app running in ``home'' mode turns off the alarm and turns on the alarm when it is in the ``away'' mode. \Soteria examines all event subscriptions and finds their corresponding event-handler methods; it creates a dummy main method for each entry point.

\begin{figure}[t!]
\begin{center}
\includegraphics[width=1\columnwidth]{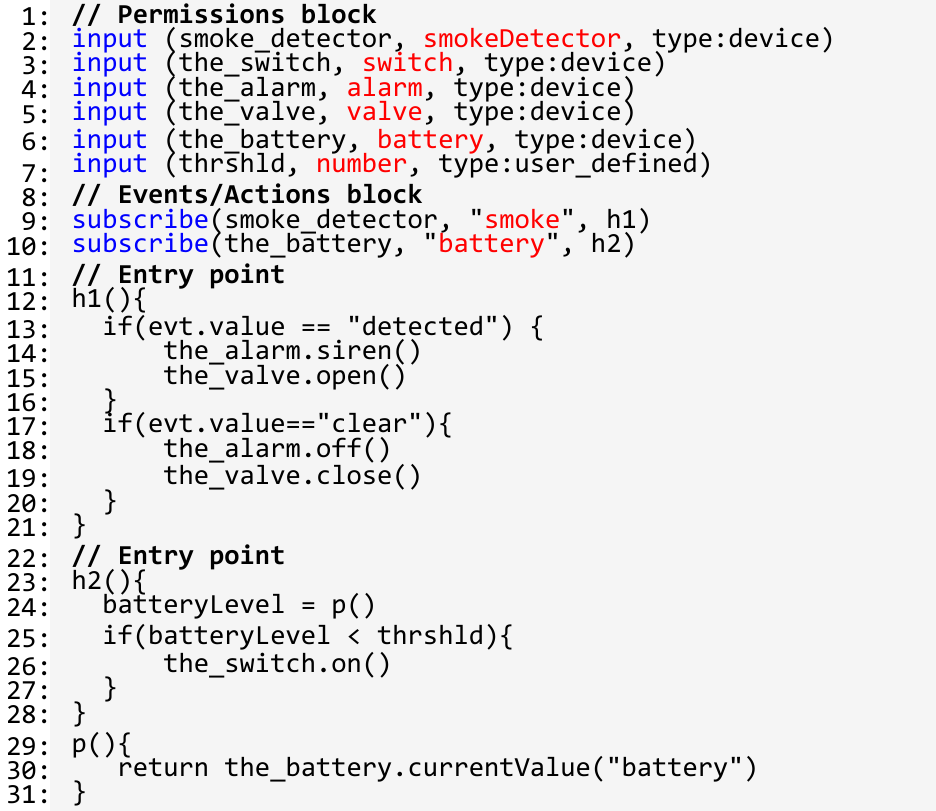}
\caption{The IR of \texttt{Smoke-Alarm} app constructed with \Soteria.}
\label{fig:IRcode}
\end{center}
\end{figure}

\vspace{2pt}\noindent\textbf{Asynchronously Executing Events.} While each event corresponds to a unique event-handler, the sequence of event handler invocations cannot be decided in advance when multiple events happen at the same time. For instance, in our example, there could be a third subscription in the event/actions block that subscribes to the switch-off event to invoke another event handler method. We consider eventually consistent events, which means any time an event handler is invoked, it will finish execution before another event is handled, and the events are handled in the order they are received by an edge device (\eg a hub). We base our implementation on path-sensitive analysis that analyzes an app's event handlers, which can run in arbitrary sequential order. This analysis is enabled by constructing a separate call graph for each entry point.

\vspace{2pt}\noindent\textbf{Call Graphs.} We create a call graph for each entry point that defines an event handler method. Turning to the IR in Fig.~\ref{fig:IRcode}, we define call graphs for two entry points \texttt{h1()} and  \texttt{h2()} (line 12 and 23). \texttt{h1()} invokes \texttt{p()} to get the current battery level of the smoke detector. 
Addressed below, note that these initial graphs are sometimes incomplete because of dynamic method invocations (reflection).

\subsection{State Model Extraction}
\label{sec:fsm-extraction}
\Soteria next extracts a state model from the IR model. 

\vspace{2pt}\noindent\textbf{Definition of State Models.} An IoT app manages one or more devices. Each device has a set of attributes, which are the states of the device. For instance, in the \texttt{Water-Leak-Detector} app, the water sensor has a boolean-typed attribute, whose value signals the ``water-detected'' or ``water-undetected'' status. Hence, we naturally model the states in the model from the values of device attributes. IoT apps are event-driven: events such as state changes or user input trigger event handlers, which can in turn change device attributes by invoking device actions. Therefore, by analyzing an IoT app's code, we can add state transitions and label them with events that trigger the transitions (changes to attribute values).

More formally, we define the state model of an IoT app as a triple $\mathtt{(Q, \Sigma, \delta)}$, where $\mathtt{Q}$ is a set of states,  $\Sigma$ is a set of transition labels, and $\mathtt{\delta}$ is a state-transition function that represents labeled transitions between states. We restrict our attention to deterministic state models, as we believe this is a condition for safe operation of IoT devices. In fact, after a state model extracted, \Soteria{} reports nondeterministic state models as a safety violation. 

\vspace{2pt}\noindent\textbf{Challenges in Extracting State Models.} Although it may appear at first glance to be straightforward, extracting state models is fraught with challenges.  First, extraction faces state-explosion problem. For instance, a thermostat device may have an integer-discrete or continuous temperature attribute would lead to many different states--adding a state for every possible value in such cases would result in state explosion. To address this, \Soteria implements a form of property abstraction that collapses states by aggregating attribute values (see Sec.~\ref{sec:model-states}).

A second challenge concerns with model precision. A state model is an abstraction of an app's logic and necessarily has to over-approximate. A sound over-approximation can cause false positives during model checking. One such approximation that caused false positives for an earlier version of \Soteria{} was that the labels on transitions were only events and thus too coarse-grained. It turns out that many IoT apps change device states \emph{conditionally}; for example, an app may turn off a switch when the energy consumption is above some threshold and turn on the switch when the energy consumption is below another threshold. For precision, the current version of \Soteria{} performs a path-sensitive analysis to extract predicates that guard state changes and adds the predicates as part of state-transition labels. We detail how state transitions are constructed in Sec.~\ref{sec:model-transitions}. 

Finally, the SmartThings platform has a number of idiosyncrasies that \Soteria{}'s model extraction must address. For instance, SmartThings apps are written in Groovy, a dynamically typed language that supports call by reflection; as another example, SmartThings apps can use special objects for persistent data storage. We will discuss how these issues are addressed in Sec.~\ref{sec:advanced-model}.

\subsubsection{Extracting States} 
\label{sec:model-states}
As discussed, states in an app's state model should represent device attribute values. Turning to the \texttt{Water-Leak-Detector} app, this app has two devices: a water sensor and a valve, both of which are represented as Boolean attributes. Therefore, the app's state model has four states: water-detected and valve-closed; water-detected and valve-open; water-undetected and valve-closed; water-undetected and valve-open. The number of possible states of an app is determined by the Cartesian product of the attributes of its device. For instance, an app implementing two devices that have $\mathtt{A}$ and $\mathtt{B}$ attributes should have states of all pairs $\mathtt{(a,b)}$, where $\mathtt{a}$$\in$$\mathtt{A}$ and $\mathtt{b}$$\in$$\mathtt{B}$.

\vspace{2pt}\noindent\textbf{Identification of Device Attributes.} An IoT platform supports many physical devices. Sound model extraction requires identifying the complete set of device attributes. Prior work has used binary instrumentation to observe the runtime behavior of apps to infer the set of device operations used with a particular state~\cite{felt2011android}. However, this is not an option on some IoT platforms such as SmartThings where app execution is inside proprietary back-ends. Another option would be to use the built-in capability files, which come with devices. The capability file for a device identifies device permissions but not attribute values--and thus do not provide enough information for analysis.

Thus, to identify device attributes, \Soteria uses platform-specific device handlers. A device handler is the representation of a physical device in an IoT platform and is responsible for communication between the device and the IoT platform (it is similar to a traditional device driver in an OS). For instance, the switch device handlers in SmartThings~\cite{smartThings-documentation} and OpenHAB~\cite{openhab} IoT platforms support the ``switch on'' and ``switch off'' attributes, and allow apps to incorporate different kinds of switches in the same way. We developed a crawler script, which visits the \texttt{status} (for attributes) and \texttt{reply} (for actions) code blocks of SmartThings device handlers found in its official GitHub repository~\cite{smartThings-documentation} and determines a complete set of attributes and actions for devices. We then created our own platform-specific \emph{device capability reference file}, which includes for each device its complete set of attributes and actions. \Soteria{} then uses this file to identify all attributes for those devices used in an app.

\begin{algorithm}[t!]
	\DontPrintSemicolon
    \footnotesize
    \setstretch{1}
	\SetKwInOut{Input}{Input}
    \SetKwInOut{Output}{Output}
	\DontPrintSemicolon
	\SetNoFillComment 
	\Input{ICFG: Inter-procedural control flow graph}
	\Input{A numerical-valued attribute}
	\Output{Dependence relation $dep$}
	$ worklist \gets \emptyset;\ done \gets \emptyset;\ dep \gets \emptyset$\;
	\For{an $id$ \normalfont{in a device action call that sets the attribute at node n}}{
		$worklist$ $\gets$ $worklist$ $\cup$ \normalfont{\{(n: $id$)\}}\;
	}
	\While{worklist \normalfont{is not empty}}{
		(\normalfont{n}: $id$) $\gets$ $worklist.pop()$\;
		$done$ $\gets$ $done$ $\cup$ \normalfont{\{(n: $id$)\}}\;
		\tcc{a def of  (n: $id$) at node n$'$ means a path from n' to n exists and on the path there is no other assignment to $id$}
		\For{\normalfont{a def of (n: $id$) at node n$'$ of form $id$ = e and e has only a single identifier $id'$}}{
			$worklist$ $\gets$ $worklist$ $\cup$ (\{(n$'$: $id'$)\} $\setminus$ $done$)\;
			$dep$ $\gets$  $dep$ $\cup$ \{(n: $id$, n$'$: $id'$) \}
		}
	}

    \caption{\footnotesize{Computing dependence from device's code}}
    \label{algo:property_abstraction}
\end{algorithm}

\vspace{2pt}\noindent\textbf{Numerical-Valued Device Attributes.} Noted above, IoT devices may have attributes with integer or continuous values leading to many states. Returning to the previous \texttt{Thermostat-Energy-Control} app, a thermostat with 45 values (50-95$\,^{\circ}\mathrm{F}$) and a power meter with 100 energy levels would lead to (clearly intractable) 4.5K states if a state is added for each combination of attribute values. 

\Soteria performs property abstraction~\cite{beyer2017combining} to reduce the state space. It first performs dependence analysis on an app's source code to identify possible sources for numerical-valued attributes, and then prunes sources using path- and context-sensitivity; the remaining sources are used to construct states in the state model. The \Soteria dependence analysis is presented in Algorithm~\ref{algo:property_abstraction} as a worklist-based algorithm. The goal of the algorithm is to identify a set of possible sources that a numerical-valued attribute can take during the execution of an app. The worklist is initialized with identifiers that are used in the arguments of device action calls that change the attribute. The worklist also labels an identifier with node information to uniquely identify the use of an identifier, because the same identifier can be used in multiple locations. The algorithm then takes an entry $(n, id)$ from the worklist and finds a definition for $id$ according to the ICFG; if the right-hand side of the definition has a single identifier, the identifier is added to the worklist;\footnote{We found that SmartThings IoT apps most often propagates a developer-defined constant or a user input to places that change device attributes. Occasionally, simple arithmetic is performed; for example, the user input is stored in $y$, followed by $x=y+10$, followed by a device attribute change using $x$. In theory, an IoT app could perform operations like $x=y+z$, where both $y$ and $z$ are user input or defined to be constants; however, we have not encountered this in our evaluation.} furthermore, the dependence between $id$ and the right-hand side identifier is recorded in $dep$. For ease of presentation, the algorithm treats parameter passing as inter-procedural definitions.

\begin{figure}[t!]
\begin{center}
\includegraphics[width=1\columnwidth]{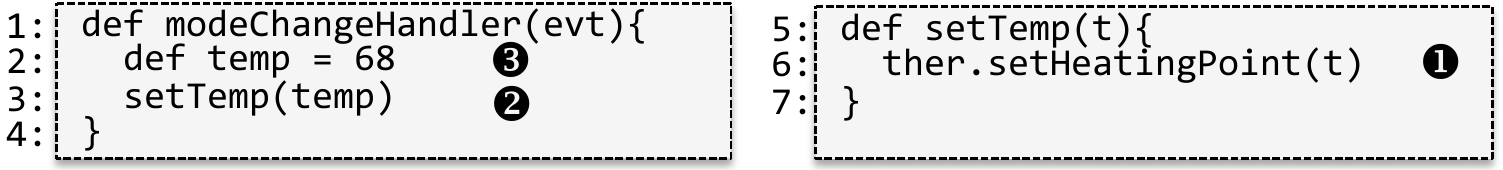}
\caption{Property abstraction under backward flow analysis.}
\label{fig:property-ther}
\end{center}
\end{figure}

The dependence analysis is a form of backward taint analysis and produces a set of sources that can affect a change to a numerical-valued attribute. For those sources, \Soteria makes them separate states in the state model and adds another state representing the rest of values.

To illustrate, we use a code block of the \texttt{Thermostat\-Energy-Control} app as an example, shown in Fig.~\ref{fig:property-ther}. There is a device action call that sets the thermostat to {\tt t} at \circled{1}; so the worklist is initialized to be (6:{\tt t}); for presentation, we use line numbers instead of node numbers to label identifiers. Then, because of the function call at \circled{2}, (3:{\tt temp}) is added to the worklist and the dependence (6:{\tt t}, 3:{\tt temp}) is recorded in $dep$. With this dependence analysis, \Soteria computes that the value for {\tt t} has to be 68$\,^{\circ}\mathrm{F}$ since {\tt temp} is initialized to be a constant value at \circled{3}. Therefore, the state model has two states for the thermostat: a state when the temperature is equal to 68$\,^{\circ}\mathrm{F}$, and a state when the temperature is not 68$\,^{\circ}\mathrm{F}$; thus, the state space for temperature values is reduced from 45 to 2. 

The backward dependence analysis also produces the $dep$ relation, through which \Soteria constructs paths from identifier initialization points to where device changes happen. For the example in Fig.~\ref{fig:property-ther}, it produces the path \circled{3}$\rightarrow$\circled{2}$\rightarrow$\circled{1}. Some produced paths by dependence analysis, however, can be infeasible paths. As an optimization, \Soteria prunes infeasible paths using path- and context-sensitivity. For a path calculated in dependence analysis, it collects the predicates at conditional branches and checks whether the conjunction of those predicates (\ie the path condition) is always false; if so, the path is infeasible and discarded. This is similar to how symbolic execution prunes paths using path conditions. For instance, if a path goes through two conditional branches and the first branch evaluates $x>1$ to true and the second evaluates $x<0$ to true, then it is an infeasible path. \Soteria does not use a general SMT solver to check path conditions. We found that the predicates used in IoT apps are extremely simple in the form of comparisons between variables and constants (such as $x=c$ and $x>c$); thus, \Soteria implemented its simple custom checker for path conditions. Furthermore, \Soteria throws away paths that do not match function calls and returns (using depth-one call-site sensitivity~\cite{SharirP81Inter}). At the end of the pruning process, we get a set of feasible paths that propagate sources defined by the developer or by user input to device action calls that change the numerical-valued attribute; and then those sources are used to define the states in the model.

\subsubsection{Extracting State Transitions}
\label{sec:model-transitions}
 If an event handler changes a device's attributes by actuating the device, it leads to a state transition. By statically analyzing event handlers, \Soteria computes state transitions and labels them with events.  When a water-detected event is generated in the \texttt{Water-Leak-Detector} app a handler method closes the valve; by analyzing the handler method, \Soteria adds a transition with the water-detected event label from state ``water-undetected and valve-open'' to ``water-detected and valve-closed'' state. 

\begin{figure}[t!]
\begin{center}
\includegraphics[width=\columnwidth]{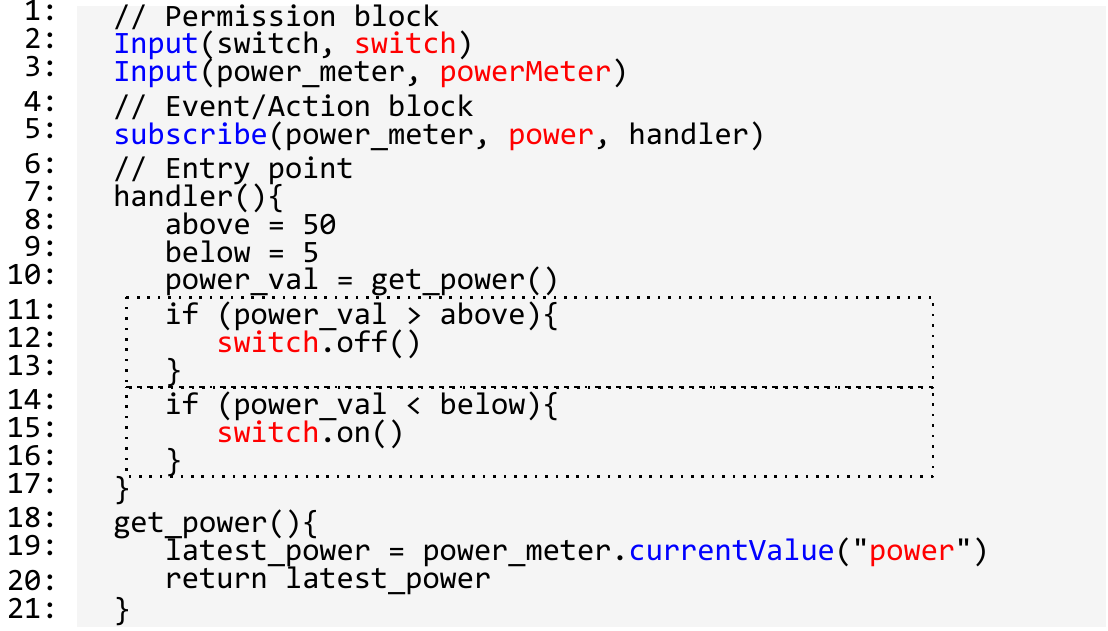}
\caption{The impact of predicates on state transitions in the \texttt{Thermostat-Energy-Control} app.}
\label{fig:path-sensitive-example}
\end{center}
\end{figure}

\vspace{2pt}\noindent\textbf{Labeling Transitions with Predicates.} Many device state changes happen in conditional branches; as a result, those state changes occur only when the predicates in the conditional branches hold. To illustrate, consider the source code in Fig.~\ref{fig:path-sensitive-example} abstracted from the \texttt{Thermostat-Energy-Control} app. The app has a conditional branch turning off the switch when energy usage is above a consumption threshold (\texttt{above}=50); it turns on the switch when it is below the threshold (\texttt{below}=5). 
 
\Soteria implements a path-sensitive analysis to capture state transitions and predicates that guard transitions. Particularly, it uses \emph{symbolic execution} to perform path exploration on source code and accumulates path conditions during exploration. In detail, it starts the analysis at the entry of an event handler with respect to some initial state, say $\mathtt{S_0}$. Then it performs forward symbolic execution along all paths, and also smartly merges paths following the ESP algorithm~\cite{das2002esp} (as a way of avoiding path explosion). For a conditional branch with condition $b$, it evaluates both paths and labels the true path with $b$ and the false path with $\lnot b$. If the end states for the true and false branches are the same, then the two paths are merged~\cite{das2002esp}. On the other hand, if the end states are different for the two paths, they are kept separate for further symbolic execution. \Soteria{} throws away infeasible paths in a way similar to that used during property abstraction. At the end of symbolic execution, \Soteria{} obtains the set of paths, their end states, and path conditions. For each path, a state transition from the initial state to the end state is added to the state model, and the transition is labeled by the event triggering the event handler and path condition.

We use the \texttt{Thermostat-Energy-Control} app with the initial state of ``switch-on'' as an illustration of this exploration. \Soteria explores all paths, and there are two feasible paths at the end, with \texttt{currentValue(``power'')>50} as the path condition of the path that turns off the switch, and \texttt{currentValue(``power'')<5} as the path condition of the path that turns on the switch.

In addition, \Soteria also tracks the sources of components in predicates that guard state transitions. For predicate \texttt{currentValue(``power'')>50} in the previous example, \texttt{currentValue(``power'')} is obtained from a device state and therefore is labeled as ``device-state'', while 50 is hardcoded by the developer and therefore is labeled as ``developer-defined''. In some cases, users can also define part of predicates at install time of an app. For instance, if the threshold value were entered by a user, then \Soteria{} would label it as ``user-defined''. Labeling sources in predicates is useful for precisely stating properties used in model checking. For example, one property says that the alarm must siren when the main door is left open longer than a threshold entered by the user. In this case, there is no property violation if the threshold is not hard-coded into the app by the developer. We detail this in Sec.~\ref{sec:property-identification}.

\begin{lstlisting}[float=t!, caption=Sample code blocks for SmartThings Idiosyncrasies, label=general-challenges, escapeinside={(*}{*)}]
/* A code block of an app using platform-specific interfaces */
subscribe(theMotion, "motion", motionHandler)
subscribe(theThermostat, "thermostat", thermostatHandler)
// different interfaces to get device attribute values
def thermostatHandler() {
    def tempAttr = theThermostat.currentState("temperature")
    def tempAttr2 = theThermostat.currentThermostat
}
// transitions without explicit event subscriptions
def motionHandler(evt) {
    if (evt.value == "active") { ... } 
    else if (evt.value == "inactive") { ... }
}
/* A code block of an app using call by reflection */
//initial state = (*\scriptsize{$\mathtt{S_0}$}*) 
def getMethod(){
  httpGet("http://url"){ resp ->
      if(resp.status == 200){
          name = resp.data.toString()
      }
  }
  "$name"() // dynamic method invocation
}           
// check state transition from (*\scriptsize{$\mathtt{S_0}$}*) to next state in both methods
def foo() {...}
def bar() {...}
/* A code block of an app using a state variable */
subscribe(theSwitch, "switch.on", turnedOnHandler)
def turnedOnHandler() {
    state.counter = state.counter + 1 
    if (state.counter>threshold){
 	// invoke device actions that lead state transitions
    }
}
\end{lstlisting}

\subsubsection{SmartThings Idiosyncrasies}
\label{sec:advanced-model}
\noindent\textbf{Platform-specific Interfaces.} The SmartThings platform implements a variety of programmer interfaces for an app to obtain device attribute values (for the same value). For instance, the temperature value of a thermostat can be read through the \texttt{currentState} or the \texttt{currentTemperature} interface (see Listing~\ref{general-challenges} (lines 1--8). Additionally, we found that some apps subscribe to all device events instead of specific device events; for example, the \texttt{subscribe} interface in Listing~\ref{general-challenges} (lines 9--13) is used to subscribe to all events of a motion sensor. The event handler then gets an event value as an argument that describes what event it is. We extract precise state models by parsing the event values passed in these interfaces and adding state transitions through those interfaces.

\vspace{2pt}\noindent\textbf{Call by Reflection.} The Groovy language supports programming by reflection (using the \texttt{GString} feature) \cite{smartThings-documentation}, which allows a method to be invoked by providing its name as a string. For instance, a Groovy method \texttt{foo()} can be invoked by declaring a string \texttt{name=``foo''} requested from an external server via the \texttt{httpGet()} interface and thereafter called by reflection through \texttt{\$name} (see Listing~\ref{general-challenges} (lines 14--26)). To handle calls by reflection, \Soteria's call graph construction adds all methods in an app as possible call targets, as a safe over-approximation. For the example in Listing~\ref{general-challenges}, \Soteria adds both \texttt{foo()} and \texttt{bar()} to the targets of the call; then it searches for state changes in each method and extracts state transitions. 

\extended{
\vspace{2pt}\noindent\textbf{Field-Sensitive Analysis of State Variables.} IoT apps can use state variables that are stored in the external storage to persist data across executions. In SmartThings, state variables are stored in either the global \texttt{state} object or the global \texttt{atomicState} object. State variables are often used in conditional branches to guard state transitions. Listing~\ref{general-challenges} (lines 27--34) presents an example app using the \texttt{state} object to store a field named \texttt{counter} to track the number of times a switch is turned on. \Soteria applies field-sensitive analysis to track the data dependencies of all fields defined in the \texttt{state} and \texttt{atomicState} objects. For example, a state transition to the switch-off state is guarded by the predicate \texttt{counter>10}. Furthermore, \Soteria{} labels state variables in predicates as ``state-variable'', indicating they are stored in external data storage.

\vspace{2pt}\noindent\textbf{Abstract Attributes and Transitions.} In SmartThings, events can be device events, which are triggered by changes to device attributes, or abstract events, which are triggered by user actions (\eg when a user clicks on an app icon) or by a pre-defined event (\eg a location mode change from away to home). Additionally, events can change abstract attributes. For instance, \texttt{setLocationMode(newMode)} sets the location mode home or away. \Soteria's IR provides a complete set of abstract attributes that an app can change, and abstract events that an app can subscribe and their corresponding event handlers. When an app subscribes an abstract event or changes abstract attribute, \Soteria creates state attributes and state transitions their event handlers induce, since handlers for abstract events can also change device and abstract attributes.
}

\subsection{Identifying IoT Properties}
\label{sec:property-identification}
As many have found in the security and safety communities, identifying the correct set of properties to validate for a given artifact is often a daunting task.  In this work and as described below, we use established techniques adapted from other domains to systematically identify a set of properties that exercise \Soteria and are representative of the real world needs of users and environments.  That being said, we acknowledge in practice that properties are often more contextual and the methods to find them are often more art than science.  Hence, we argue that many environments will need to tailor their property discovery process to their specific security and safety needs. 

\begin{figure}[t!]
\begin{center}
\includegraphics[width=1\columnwidth]{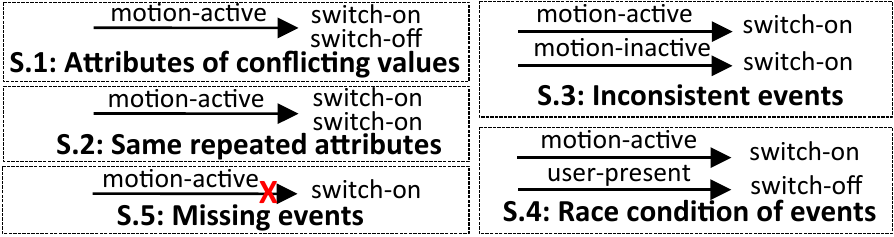}
\caption{Illustration of general properties ($\mathtt{S.1}$-$\mathtt{S.5}$). Description of properties is available in the Appendix ~\ref{appendix:properties}.}
\label{fig:s_properties}
\end{center}
\end{figure}

We refer to a property as a system artifact that can be formally expressed via specification and validated on the application model. We extend the use/misuse case requirements engineering~\cite{oracle-security, mead2007compare, schumacher2013security, yoshioka2008survey} to identify IoT properties. This approach derives requirements (properties) by evaluating the connections between 1) {\it assets} are artifacts that someone places value upon, \eg a garage door, 2) {\it functional requirements} define how a system is supposed to operate in normal environment, \eg when a garage door button is opened, the door opens,  and 3) {\it functional constraints} restrict the use or operation of assets, \eg the door must open only when an authorized garage-door opener device requests it. We used use/misuse case requirements engineering as a property discovery process on the IoT apps used in our evaluation (See Section~\ref{sec:evaluation}) and identified 5 general properties ($\mathtt{S.1}$-$\mathtt{S.5}$, see Fig.~\ref{fig:s_properties}) and  30 application-specific properties ($\mathtt{P.1}$-$\mathtt{P.30}$, see Table~\ref{table:app-specific-properties}). 

\vspace{2pt}\noindent\textbf{General Properties.} General properties are constraints on state models that are independent of an app's semantics--intuitively, these are states and transitions that should never occur regardless of the app domain. We develop the properties based on the constraints on states and state transitions. To illustrate, property $\mathtt{S.1}$ states that a handler must not change an attribute to conflicting values on the same control-flow path, \eg the motion-active handler must not turn on and turn off a switch in the same branch of the handler. More subtly, property $\mathtt{S.4}$ states that two or more non-complementary handlers must not change an attribute to conflicting values, \eg a user-present handler turns on the switch while a timer turns off the switch--leading to a potential race condition.

\vspace{2pt}\noindent\textbf{App-specific Properties.} App-specific properties are developed according to use cases of one or more devices--here we take a device-centric approach. For instance, $\mathtt{P.1}$ says that the door must always be locked when the user is not at home (thus involving the smart door and presence detector). Similarly, $\mathtt{P.30}$, states that the water valve must be shut off when there is a water leak (thus involving the water valve and moisture sensor). We evaluated all apps using this approach, but defer discussion to the extended paper. We check the app against a property if all of the devices in the property are included in the app.

\begin{table}[t!]
\newcolumntype{P}[1]{>{\centering\arraybackslash}p{#1}}
\centering
\def\arraystretch{1}
\resizebox{\columnwidth}{!}{
\begin{tabular}{|P{0.6cm}|p{9.5cm}|}
\hline
\multicolumn{1}{|c|}{\textbf{ID}} & \multicolumn{1}{c|}{\textbf{Property Description}} \\ \hline \hline
$\mathtt{P.1}$  & The door must always be locked when the user is not home. \\ \hline
$\mathtt{P.10}$ & The alarm must always go off when there is smoke. \\ \hline
$\mathtt{P.12}$ & The light must be off when the user is not home. \\ \hline
$\mathtt{P.13}$ & The devices (\eg coffee machine, crock-pot) must always \\
                & \vspace{-8pt} be on at the time set by the user.\\ \hline
$\mathtt{P.14}$ & The refrigerator and security system must always be on. \\ \hline
$\mathtt{P.17}$ & The AC and heater must not be on at the same time. \\ \hline
$\mathtt{P.22}$ & The battery of devices must not be below a specified threshold. \\ \hline
$\mathtt{P.28}$ & The sound system must not play music during the sleeping mode. \\ \hline
$\mathtt{P.29}$ & The flood sensor must always notify the user when there is water. \\ \hline
$\mathtt{P.30}$ & The water valve must be closed if a leak is detected. \\ \hline 
\end{tabular}%
}
\caption{Examples of application-specific properties.  A complete list of properties is available in the Appendix ~\ref{appendix:properties}.}
\label{table:app-specific-properties}
\end{table}

\subsection{Validating Properties}
\label{sec:property-validation}
Validation begins by defining a temporal formula for each property to be verified. Thereafter, \Soteria  uses a general purpose model checker to validate the property with respect to the generated model of the target app (see next section for details).  What the user does with a discovered violation is outside the scope of \Soteria.  However, in most cases, we expect that the results will be recorded and the code hand-investigated to determine the cause(s) of the violation. If the violation is not acceptable for the domain or environment, the app can be rejected (from the market) or modified (by the developer) as needs dictate.

Validation of properties in multi-app environments is more challenging. Apps often interact through a common device or some common abstract event (such as the home or away modes). For illustration, consider two apps (\texttt{App1} and \texttt{App2}) co-resident with the \texttt{Smoke-Alarm} and \texttt{Thermostat-Energy-Control} apps in a multi-device environment. \texttt{App1} changes the mode from away to home when the light switch is turned on, and \texttt{App2} turns off a light switch when the smoke is detected, as follows:

{\footnotesize{
\begin{nstabbing}
\textbf{\texttt{Smoke-Alarm}}: switch-off$\xrightarrow{\text{smoke-detected}}$switch-on\\[-2pt]
\textbf{\texttt{App1}}: away-mode$\xrightarrow{\text{switch-on}}$home-mode\\[-2pt]
\textbf{\texttt{Thermostat-Energy-Control}}: door-unlocked$\xrightarrow{\text{home-mode}}$door-locked \\[-2pt]
\textbf{\texttt{App2}}: switch-on$\xrightarrow{\text{smoke-detected}}$switch-off
\end{nstabbing}}}
\noindent The \texttt{Smoke-Alarm} app interacts with \texttt{App1} through the switch, and interacts with \texttt{App2} through the smoke detector and switch. The \texttt{Thermostat-Energy-Control} app interacts with \texttt{App2} through the mode-change event.

To check general and app-specific properties in the setting of multiple apps, \Soteria builds a state model that is the union of the apps' state models. The resulting state model $\mathtt{G'}$ represents the complete behavior when running the multiple apps together. The union algorithm is presented in Algorithm~\ref{algo:union-algorithm}. \Soteria first creates an empty-transition state model $\mathtt{G'}$ whose states are the Cartesian product of the states in the input apps (line 1); note that since the input apps' states encode device attributes, the Cartesian product should remove attributes of duplicate devices (i.e., those devices that appear in multiple apps). For instance, if we consider \texttt{Smoke-Alarm} and \texttt{App1}, $\mathtt{G'}$ should have four states, and each state encodes a pair of switch and mode attributes. The algorithm then iterates through all apps' transitions and adds appropriate transitions to the union model $\mathtt{G'}$. \Soteria's union algorithm is a modification of the multiple-graph union algorithm of igraph library~\cite{igraph}, based on a set of constraints on transitions and states. It has a complexity of $\mathtt{O(|V|+|E|)}$, $\mathtt{|V|}$ and $\mathtt{|E|}$ is the number of vertices and edges in $\mathtt{G'}$. 

\begin{algorithm}[t]
	\DontPrintSemicolon
	\setstretch{1}
    \footnotesize
	\SetKwInOut{Input}{Input}
    \SetKwInOut{Output}{Output}
	\DontPrintSemicolon
	\SetNoFillComment 
	\Input{$\mathtt{G}=\mathtt{\{G_i\}_{i=1}^{n}}$: State models of $n$ apps}
	\Output{$\mathtt{G'}$ is the union of $\mathtt{\{G_i\}_{i=1}^{n}}$}
	\tcc{Initialize $\mathtt{G'}$}
	$\mathrm{states}(\mathtt{G'})$ $\gets$ $\{v\ |\ v\ \mbox{is a tuple of attribute values in}\ \mathtt{G} \}$\;
	\tcc{Construct union of apps' state models}
	\For{i $\in$ (1: n)}{
		\ForAll{states $v$ $\in$ $\mathtt{G}_i$}{
			\ForAll{transitions $e= v \xrightarrow{l} u$ $\in$ $\mathtt{G}_i$}{	
				$\mathtt{V'}$ is a subset of states in $\mathtt{G'}$ that contain $v$\;
				$\mathtt{U'}$ is a subset of states in $\mathtt{G'}$ that contain $u$\;
				\ForAll{$v' \in \mathtt{V'}$ \normalfont{and} $u' \in \mathtt{U'}$}{
					add $e'=v'\xrightarrow{l} u'$ to $\mathtt{G'}$ and label the edge with $i$\;
				}
			}
		}
	}
    \caption{\footnotesize{Creating the union of apps' state models}}
    \label{algo:union-algorithm}
\end{algorithm}

With the union state model created, \Soteria{} then performs model checking on the union model concerning properties we discussed earlier. As an example, \Soteria reports that, when \texttt{Smoke-Alarm} and \texttt{App2} are used together, there is a property violation of  $\mathtt{S.1}$: the smoke-detected event would make the \texttt{Smoke-Alarm} app turn on the switch, while it would also make \texttt{App2} to turn off the switch. As another example, when \texttt{Smoke-Alarm}, \texttt{App1} and \texttt{Thermostat-Energy-Control} are used together, there is a misuse case that violates property $\mathtt{P.3}$: the door would be locked when there is smoke at home. The property violation is demonstrated as follows:

\vspace{2pt}

{\scriptsize{
\begin{nstabbing}
\textbf{
switch-off$\xrightarrow{\text{smoke-detected}}$switch-on $\xrightarrow{\text{switch-on}}$home-mode$\xrightarrow{\text{home-mode}}$door-locked}
\end{nstabbing}}}

\noindent $\mathtt{P.3}$ is violated because switch-on attribute in the \texttt{Smoke-Alarm} app is used by \texttt{App1}, which changes the mode from away to home. The mode change then triggers locking the door in \texttt{Thermostat-Energy-Control}.

\begin{figure*}[!th]
    \centering{\includegraphics[width=\textwidth]{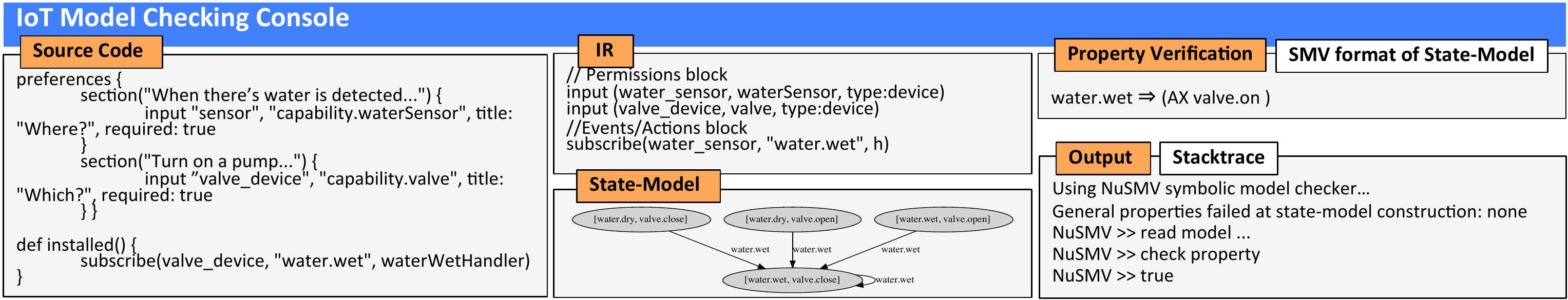}}
    \caption{Our \Soteria framework designed for IoT apps. The left region is the analysis frame; the middle region contains the IR and visual representation of the state model of an example IoT app, and the right region shows the output for a property violation.}
    \label{fig:webapp}
\end{figure*}

\extended{
\begin{figure}[t!]
\begin{center}
\includegraphics[width=1\columnwidth]{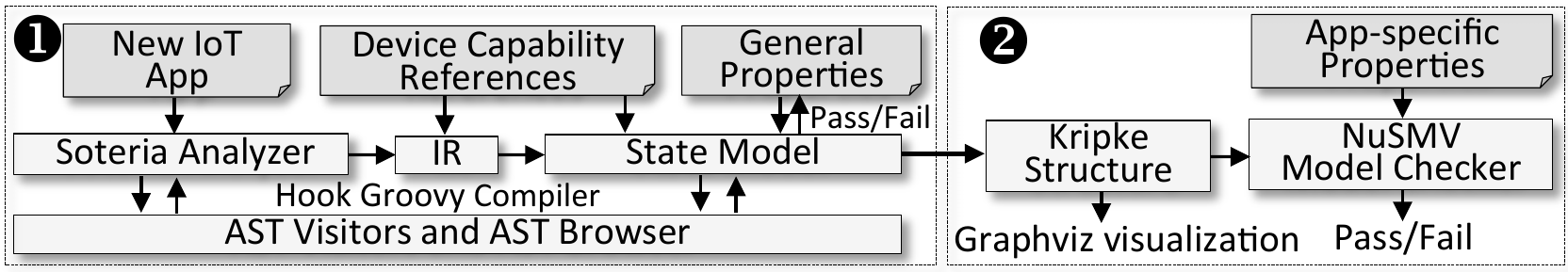}
\caption{\Soteria's implementation on SmartThings.}
\label{fig:implementation}
\end{center}
\end{figure}
}

\section{Implementation} 
\label{sec:implementation}
\noindent\textbf{IR and State Model Construction.} Constructing an IR from the source code requires, among other things, the building of the app's ICFG. Here the \Soteria IR-building algorithm directly works on the Abstract Syntax Tree (AST) representation of Groovy source code. The Groovy compiler supports customizing the compilation via compiler hooks, through which one can insert extra passes into the compiler (similar to the modular design of the LLVM compiler~\cite{llvm}). \Soteria visits AST nodes at the compiler's semantic analysis phase where the Groovy compiler performs consistency and validity checks on the AST. Our implementation uses an \texttt{ASTTransformation} to hook into the compiler, \texttt{GroovyClassVisitor} to extract the entry points and the structure of the analyzed app, and \texttt{GroovyCodeVisitor} to extract method calls and expressions inside AST nodes. Here we AST visitors to analyze expressions and statements to construct the IR and model. 

\Soteria uses AST visitors for state model construction as well. We extend the \texttt{ASTBrowser} class implemented in the Groovy Swing console, which allows users to enter and run Groovy scripts~\cite{groovy-console}. The implementation hooks into the IR of an app in the console and dumps information to the \texttt{TreeNodeMaker} class; the information includes an AST node's children, parent, and all properties built during compilation. This includes the resolved classes, static imports, the scope of variables, method calls, interfaces accessed in an app. We then use Groovy visitors to traverse the IR's ICFG and extract the state model.

\extended{
Since Groovy is a JVM-hosted language, one natural approach would be first to compile Groovy code into Java bytecode using the Groovy compiler and then build the IR via the help of the Soot analysis framework~\cite{vallee1999soot}. However, this approach was not feasible due to the heavy use of reflection in the bytecode generated by the Groovy compiler. In particular, the Groovy compiler translates every direct method call into a call by reflection. For instance, the \texttt{Smoke-Alarm} app in Fig.~\ref{fig:IRcode} is compiled to bytecode with 14 reflective calls. Soot, unfortunately, does not produce good analysis results when the input bytecode uses reflection, as our experience finds. 
}

\vspace{2pt}\noindent\textbf{Model Checking with NuSMV.} We translate the state model of an IoT app into a Kripke structure~\cite{clarke1999model}. A Kripke structure is an equivalent temporal structure of a state model and increases readability. We create a visual representation of a state model using open-source graph visualization software GraphViz~\cite{ellson2001graphviz}. We use the open-source symbolic model checker NuSMV~\cite{cimatti2002nusmv} for its reliability and maturity. We express properties with temporal logic formulas~\cite{clarke1981design}. NuSMV either confirms a property holds or presents a counter-example showing why the property is false. To address state explosion in apps that control a large number of devices or that have complex control logic, we use NuSMV options that combine binary decision diagrams(BDDs)-based model checking with SAT-based model checking~\cite{biere1999symbolic}. This was successfully applied to verify models having more than $10^{20}$ states and hundreds of state variables~\cite{burch1991symbolic}.

\vspace{2pt}\noindent\textbf{Output of \Soteria.} Fig.~\ref{fig:webapp} presents \Soteria's analysis result on a sample app. It builds the app IR, extracts the state model, and displays a visual representation of the state model. For each property, \Soteria{} either shows the property holds or presents a counter-example.  

\section{Evaluation}
\label{sec:evaluation}
As a means of evaluating the \Soteria framework, we performed an analysis on two large-scale data-sets--one market based and one synthetic. In these studies, we sought to validate the correctness, completeness, and performance of property analysis on the target datasets. We performed our experiments on a laptop computer with a 2.6GHz 2-core Intel i5 processor and 8GB RAM, using Oracle's Java Runtime version 1.8 (64 bit) in its default settings. We use NuSMV 2.6.0 for model checking and Graphviz 2.36 for visualization of a state model. 

\vspace{2pt}\noindent\textbf{Datasets} For the market dataset, we obtained 35 official (vetted) apps ($\mathtt{O1}$-$\mathtt{O35}$) from the SmartThings GitHub repository~\cite{Official} and 30 community-contributed third-party (non-vetted) apps ($\mathtt{TP1}$-$\mathtt{TP30}$) from the official SmartThings community forum~\cite{Community} in late 2017 (see Table~\ref{table:app-properties}). The  65 apps were selected to include various devices and functionality that encompass diverse real-life use-cases. 

For the synthetic dataset, we introduce \MalIoT~\cite{iotbench}, an open source repository containing flawed IoT apps. Inspired by other security-relevant app test suites~\cite{ArztFlowdroid, EnckTaintDroid, mclaughlin2012sabot}, \MalIoT includes 17 hand-crafted flawed SmartThings apps ($\mathtt{App1}$-$\mathtt{App17}$) containing property violations in an individual app and multi-app environments. 14 apps have a single property violation, and three have multiple property violations, with a total of 20 property violations. We present the apps and their violations in Appendix~\ref{appendix:iotmalicious_apps}. The apps include various devices covering diverse real-life use-cases. The accurate identification of property violations requires program analysis including multiple entry points, numerical-valued device attributes, and transitions guarded by predicates. Each app in \MalIoT also comes with ground truth of what properties are violated; this is provided in a comment block in the app's source code.

\begin{table}[t!]
\centering
\def\arraystretch{1.1}
\setlength{\tabcolsep}{2pt}
\resizebox{\columnwidth}{!}{%
\begin{threeparttable}[b]
\begin{tabular}{l|c|c|c|c|c|}
\cline{2-6}
 & Nr. & Unique Devices & Avg/Max States\textdaggerdbl & Avg/Max LOC & Func.\textdagger \\ \hline\hline
\multicolumn{1}{|l|}{\textbf{Official}} & 35 & 14 & 36/180 & 220/2633 & All \\ \hline
\multicolumn{1}{|l|}{\textbf{Third-party}} & 30 & 18 & 32/96 & 246/1360 & All \\ \hline
\end{tabular}%
\begin{tablenotes}
	\item [\textdaggerdbl]This is after applying \Soteria's state-reduction algorithms.
	\item [\textdagger]The apps cover all spectrum of functionality, including security and safety, green living, convenience, home automation, and personal care. We determined an app's functionality by checking definition blocks in its source code.
\end{tablenotes}
\end{threeparttable}
}
\caption{Description of analyzed official and third-party apps.}
\label{table:app-properties}
\end{table}

\subsection{Market App Evaluation}
\label{sec:market_evaluation}
We first report results of the verification of general ($\mathtt{S.1}$-$\mathtt{S.5}$) and app-specific ($\mathtt{P.1}$-$\mathtt{P.30}$) properties. The properties are checked for each app and collections of apps working in concert. \Soteria flagged that nine individual apps and three multi-app groups violate at least one property. We manually checked the property violations and verified that all reported ones are true positives. The manual checking process was straightforward to perform since SmartThings apps are relatively small.

\vspace{2pt}\noindent\textbf{Individual App Analysis.} Table~\ref{table:singleappResults} the results of our analysis on single apps. \Soteria flagged one third-party app violating multiple properties, eight third-party apps violating a single property. None of the official apps were flagged as violating properties; we believe this is because of the strict manual vetting enforced on official apps, which takes a couple of months~\cite{smartThings-review}. For third-party apps, we manually verified that all reported property violations are indeed problems with the implementation. For example, a property violation happens in an app ($\mathtt{TP6}$) that turns off and on a light switch when there is nobody at home; another app ($\mathtt{TP9}$) unlocks the door at sunset and locks the door at sunrise--and unintended action.

\begin{table}[t!]
\newcolumntype{P}[1]{>{\centering\arraybackslash}p{#1}}
\centering
\def\arraystretch{0.9}
\resizebox{\columnwidth}{!}{%
\begin{tabular}{|P{0.7cm}|p{9.5cm}|P{1.8cm}|}
\hline
\textbf{ID} & \multicolumn{1}{|c|}{\textbf{Violation Description}} & \multicolumn{1}{c|}{\textbf{Violated Pr.}} \\ \hline \hline
$\mathtt{TP1}$ & The music player is turned on when user is not at home. & $\mathtt{P.13}$ \\ 
$\mathtt{TP2}$ & The switch turns on and blinks lights when no user is present. & $\mathtt{P.12}$ \\ 
$\mathtt{TP3}$ & The location is changed to the different modes when the switch & $\mathtt{S.4}$ \\
& is turned off and when the motion is inactive.  & \\
$\mathtt{TP4}$ & The flood sensor sounds alarm when there is no water. & $\mathtt{P.29}$ \\ 
$\mathtt{TP5}$ & The music player turns on when the user is sleeping. & $\mathtt{P.28}$ \\ 
$\mathtt{TP6}$ & The lights turn on and turn off when nobody is at home. & $\mathtt{P.13}$ and $\mathtt{S.1}$ \\ 
$\mathtt{TP7}$ & The lights turn on and turn off when the icon of the app is tapped. & $\mathtt{S.1}$ \\ 
$\mathtt{TP8}$ & The door is unlocked on sunrise and locked on sunset. & $\mathtt{P.1}$ \\ 
$\mathtt{TP9}$& The door is locked multiple times after it is closed. & $\mathtt{S.2}$ \\ \hline
\end{tabular}
}
\caption{\Soteria's results on individual apps.}
\label{table:singleappResults}
\end{table}

To assess whether the property violations are real bugs in analyzed apps, we opened a thread in official SmartThings community forum and asked users whether the functionality of the apps confirms their expectations~\cite{smartThingsFromQuestion}. We got eight answers from the users that are smart home enthusiasts. These apps may have subtle and surprising uses under the right conditions: a user for $\mathtt{TP4}$, said that he used his flood sensor to let him know when there is no water so that he can add water to the trees during Christmas; another user stated that $\mathtt{TP6}$ might simulate occupancy of his home at night by randomly turning on/off lights when nobody is home. To guard against malicious code, those users stated that they attempted to read and understand the source code of the apps before they installed them. However, since regular users cannot be expected to read and check the source code of apps manually, \Soteria addresses this problem by analyzing apps and presenting their potential property violations to users, which allows them to determine whether a violation is actually harmful.

\vspace{2pt}\noindent\textbf{Multi-App Analysis.} We found that multiple apps working in concert can lead to unsafe and undesired device states. \Soteria flagged three group of apps violating multiple properties. We examined 28 groups and found three groups that have 17 apps violate 11 properties. Table~\ref{table:multi-app_results} shows the app groups, events, and device attributes that constitute violations, and violated properties. In the following discussion, we will use app group IDs ($\mathtt{G.1}$-$\mathtt{G.3}$) in Table~\ref{table:multi-app_results}. Each group includes a set of apps that a user may install together and authorize to use the same devices. 

\begin{table}[t!]
\centering
\def\arraystretch{0.7}
\resizebox{\columnwidth}{!}{%
\begin{tabular}{|c|c|l|c|}
\hline
\textbf{Gr. ID} & \multicolumn{1}{c|}{\textbf{App ID}} & \multicolumn{1}{c|}{\textbf{Event/Actions}} & \textbf{Violated Pr.} \\ \hline  \hline
\multirow{4}{*}{$\mathtt{G.1}$} & $\mathtt{O3}$ & $\xrightarrow{\text{contact sensor open}}$switch on & 
\multirow{4}{*}{\begin{tabular}[c]{@{}c@{}}\\ $\mathtt{S.1}$, $\mathtt{S.2}$, \\ $\mathtt{S.3}$\end{tabular}}\\ \cline{2-3}
 & \multirow{2}{*}{$\mathtt{O4}$} & $\xrightarrow{\text{contact sensor open}}$switch off &  \\ \cline{3-3}
 &  & $\xrightarrow{\text{contact sensor close}}$switch on &  \\ \cline{2-3}
 & $\mathtt{O8}$, $\mathtt{TP12}$ & $\xrightarrow{\text{contact sensor close}}$switch off &  \\ \hline
\multirow{4}{*}{$\mathtt{G.2}$}& $\mathtt{O14}$ & $\xrightarrow{\text{contact sensor open}}$switch off & 
\multirow{4}{*}{\begin{tabular}[c]{@{}c@{}} $\mathtt{S.2}$, $\mathtt{S.4}$\end{tabular}}\\ \cline{2-3}
 & $\mathtt{O9}$, $\mathtt{O16}$, $\mathtt{TP3}$ & $\xrightarrow{\text{motion active}}$switch on &  \\ \cline{2-3}
 & $\mathtt{TP2}$  & $\xrightarrow{\text{app touch}}$switch on &  \\ \hline
\multirow{6}{*}{$\mathtt{G.3}$} & \multirow{2}{*}{$\mathtt{O7}$, $\mathtt{TP3}$} & $\xrightarrow{\text{switch off}}$change location mode & \multirow{6}{*}{\begin{tabular}[c]{@{}c@{}}\\ $\mathtt{P.12}$, $\mathtt{P.13}$, \\ $\mathtt{P.14}$, $\mathtt{P.17}$,\\ $\mathtt{S.1}$, $\mathtt{S.2}$  \end{tabular}} \\ \cline{3-3}
 &  & $\xrightarrow{\text{motion inactive}}$ change location mode &  \\ \cline{2-3}
 &  $\mathtt{O30}$, $\mathtt{TP21}$ & $\xrightarrow{\text{location mode change}}$switch off &  \\ \cline{2-3}
 &  $\mathtt{O31}$, $\mathtt{TP22}$ & $\xrightarrow{\text{location mode change}}$switch on &  \\ \cline{2-3}
 & \multirow{2}{*}{$\mathtt{O12}$, $\mathtt{TP19}$} & $\xrightarrow{\text{location mode change}}$set thermostat heating &  \\ \cline{3-3}
 &  & $\xrightarrow{\text{location mode change}}$set thermostat cooling &  \\ \hline 
\end{tabular}
}
\caption{\Soteria's results in multi-app environments.}
\label{table:multi-app_results}
\end{table}

In $\mathtt{G.1}$, $\mathtt{O3}$ and $\mathtt{O4}$ violate $\mathtt{S.1}$ by setting the switch attribute to conflicting values when the contact sensor is open; there is a similar violation between $\mathtt{O4}$, $\mathtt{O8}$ and $\mathtt{TP12}$ when the contact sensor is closed. $\mathtt{O8}$ and $\mathtt{TP12}$ violates $\mathtt{S.2}$ by turning on the switch multiple times with the ``contact sensor close'' event. In addition, $\mathtt{O3}$ and $\mathtt{O4}$ violate $\mathtt{S.3}$ by turning on the switch with complement events of ``contact sensor close'' and ``contact sensor open''. 
In $\mathtt{G.2}$, $\mathtt{O9}$, $\mathtt{O16}$, and $\mathtt{TP3}$ violates $\mathtt{S.2}$ by turning on the switch multiple times with the ``motion active'' event. Additionally, the interaction between $\mathtt{O14}$, $\mathtt{O9}$, $\mathtt{O16}$ and $\mathtt{TP3}$ violates $\mathtt{S.4}$ by invoking ``switch on'' and ``switch off'' actions with different device events (``contact sensor open'' and ``motion active''). There is a similar violation between $\mathtt{O14}$ and $\mathtt{TP2}$  (``contact sensor open'' and ``app touch''). These events may occur at the same time, which leads to a race condition. 
In $\mathtt{G.3}$, similar to the other groups, $\mathtt{S.1}$ and $\mathtt{S.2}$ are violated. In addition, multiple app-specific properties are violated. $\mathtt{O7}$ and $\mathtt{TP3}$ change the location mode when the switch is turned off and also when motion is inactive. $\mathtt{O30}$ and $\mathtt{TP21}$ turn off the switch of a set of devices including a security system, smoke detector, and heater when the location is changed; $\mathtt{O31}$ and $\mathtt{TP22}$ turns on devices such as TV, coffee machine, A/C, and heater when the location is changed; both cases violate multiple properties ($\mathtt{P.12}$, $\mathtt{P.13}$, $\mathtt{P.14}$ and $\mathtt{P.17}$) and cause security and safety risks for users. Lastly, $\mathtt{O12}$ and $\mathtt{TP19}$ sets the thermostat to user settings when the switched is turned off and when the motion is inactive. These result in an unauthorized control of thermostat heating and cooling temperature values. 

\subsection{\textsc{MalIoT} Evaluation}
\label{sec:iotmalicious}
Our analysis of \Soteria on \MalIoT showed that it correctly identified the 17 of the 20 unique property violations in the 17 apps. \Soteria produces a false warning for an app that uses call by reflection ($\mathtt{App5}$). This app invokes a method via a string. It over-approximates the call graph by allowing the method invocation to target all methods in the app. Since one of the methods turns off the alarm when there is smoke, \Soteria reports a violation. However, it turns out that the reflective call in this app would not call the property-violating method. Note this pattern did not appear in the 65 real IoT apps we discussed earlier. Additionally, \Soteria did not report a violation for an app that leaks sensitive data ($\mathtt{App10}$) and for an app that implements dynamic device permissions ($\mathtt{App11}$) as they are outside the scope of \Soteria analysis.

\begin{figure}[t!]
\begin{center}
\includegraphics[width=1\columnwidth]{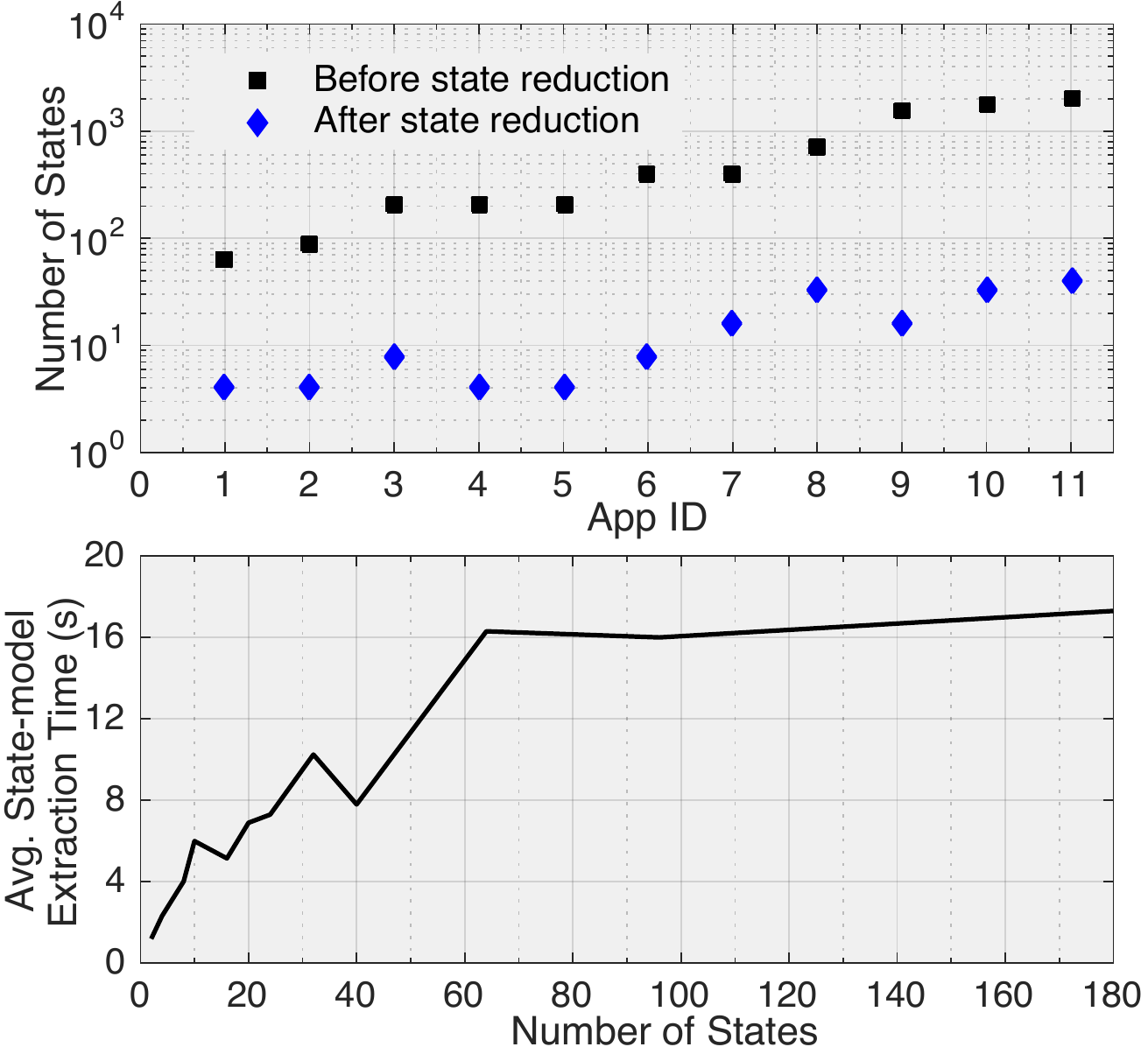}
\caption{\Soteria's state reduction efficacy (Top). \Soteria's state model extraction overhead (Bottom).}
\label{fig:performance}
\end{center}
\end{figure}

\subsection{MicroBenchmarks}
\label{sec:microbenchmarks}
\noindent\textbf{State Reduction Efficacy.} Earlier we presented algorithms for performing property abstraction on numerical-valued device attributes. To evaluate its impact, we measured the number of states before and after the application of these algorithms, and the results are presented on the top of Fig.~\ref{fig:performance}. We note that \Soteria{} performs state reduction only for apps with devices that have numerical-valued attributes; examples include thermostats, batteries, and power meters. Among the devices we examine, there are ten such devices in analyzed apps, and 14 apps grant access to these devices, and the states of three apps have the same number before reduction and reduced to the same number. The figure shows that \Soteria{}'s state reduction often results in order of magnitude less number of states.

\vspace{2pt}\noindent\textbf{State Model Extraction Overhead.} We ran \Soteria with apps that have varying numbers of states and recorded the state-model generation time; the result is shown on the bottom of Fig.~\ref{fig:performance}. The time includes the time for IR extraction, generating a graphical representation of the model, obtaining the SMV code of a state model, and logging (required for general properties). The average run-time for an app with 180 states was 17.3$\pm$2 secs. We note that the total time depends on the time taken by the algorithms we have developed for state reduction. For instance, an app having 32 states took more time than an app having 40 states due to many branches used in the 32-state app. Note that overheads can be mitigated by eliminating non-essential processing and other optimization. 

We also measured the time for constructing a state model in multi-app environments. The state model of multiple apps requires extraction of each app's state model. \Soteria's graph-union algorithm then finds 30 interacting apps (which have on average 64 states and six state attributes) and 4$\pm$2.1 seconds for the union algorithm.

\vspace{2pt}\noindent\textbf{Property Verification Overhead.} We evaluated the verification time of a property on state models. The verification of a property took on the order of milliseconds to perform since the SmartThings apps have comparatively smaller state models than the large-scale ones found in other domains such as operating system kernels. 

\section{Limitations and Discussion}
\label{sec:limit-discuss}
A limitation of \Soteria is the treatment of call by reflection. As discussed in Sec.~\ref{sec:advanced-model}, \Soteria constructs an imprecise call graph that allows a reflective call to target any method. This increases the size of state models and may lead to false positives during property checking. We plan to explore string analysis to statically identify possible values of strings and refine the target sets of method calls by reflection. Another limitation of \Soteria is dynamic device permissions and app configurations. These may yield property violations because of the erroneous device and input configurations by users at install time. For instance, if a user enters an incorrect time value, the door may be left unlocked in the middle of the night.

\Soteria's implementation and evaluation are based on the SmartThings programming platform designed for home automation. There are other IoT domains suitable for applying model checking for finding property violations, such as FarmBeats for agriculture~\cite{farmBeats}, HealthSaaS for healthcare~\cite{healthIoT}, and KaaIoT for the automobile industry~\cite{Kaa}. We plan to extend our \Soteria to these platforms by applying the IR-based analysis, as well as engage in large-scale analyses of IoT markets and industries.

\extended{
\section{Related Work}
There has been an increasing amount of recent research exploring IoT security and more broadly safety. These systems centered on the security of emerging IoT programming platforms and IoT devices~\cite{fernandes2016security, xu2014security}. Other efforts have explored vulnerability analysis within specific IoT devices such as light systems~\cite{oluwafemi2013experimental} and smart locks~\cite{ho2016smart}. These works have found that apps can be easily exploited to gain unauthorized access to control devices. Several efforts have also focused on sensitive data leaks~\cite{saint-taint-analysis} and correctness of IoT apps using a range of analyses~\cite{fernandes2016flowfence, rahmati2016applying, jia2017contexiot}. In contrast, we build the first system that addresses IoT-specific challenges and also addresses platform- and language-specific problems for checking general- and app-specific properties in individual apps or multiple interacting apps.

Model checking is used as a vehicle to analyze the correctness of software in security-critical systems such as investigating network vulnerabilities~\cite{ritchey2000using} and verifying Android apps against privacy sensitive data leaks~\cite{bai2017towards}. Other previous works have also attempted to model the implementation of diverse software systems to explore their state spaces~\cite{bai2014trustfound, morse2015model, yang2006using}. In contrast, to our best knowledge, we present the first system to automatically extract state models from the source code of an IoT app IR and check the correctness of an IoT app with the desired properties.
}

\section{Conclusions}
We presented \Soteria, a novel system that extracts state models from IoT code suitable for finding the security, safety, and functional errors.We evaluated \Soteria in two studies; a study of apps on the SmartThings market, and a study on our novel \MalIoT app corpus. These studies demonstrated that our approach can efficiently identify property violations and that many apps violate properties when used in isolation and when used together in multi-app environments. In future work, we will extend the kinds of analysis and provide a suite of tools for developers and researchers to evaluate implementations and study the complex interactions between users and IoT environments devices that they use to enhance their lives.


\anonymous{
\section{Acknowledgements}
The authors thank Ashutosh Pattnaik and Prasanna Rengasamy for helpful discussions about this work. We also thank Megan McDaniel for taking care of our diet before the paper deadline. Research was supported in part by the Army Research Laboratory, under Cooperative Agreement Number W911NF-13-2-0045 (ARL Cyber Security CRA) and the National Science Foundation Grant No. CNS-1564105. The views and conclusions contained in this document are those of the authors and should not be interpreted as representing the official policies, either expressed or implied, of the Army Research Laboratory or the U.S. Government. The U.S. Government is authorized to reproduce and distribute reprints for Government purposes notwithstanding any copyright notation here on. 
}

{\footnotesize 
\bibliographystyle{acm}
\bibliography{main.bbl}
}

\extended{
\appendix{}

\setcounter{figure}{0}
\setcounter{table}{0}
\setcounter{equation}{0}
\setcounter{lstlisting}{0}


\section{Source Code and IR of Example Apps}
\label{appendix:example-app}
We present the Groovy source code and Intermediate Representation (IR) of \texttt{Smoke-Alarm}, \texttt{Water-Leak-Detector}, and \texttt{Thermostat-Energy-Control} apps in Section~\ref{sec:smoke}, ~\ref{sec:water}, and ~\ref{sec:thermostat}, respectively. 

\subsection{Smoke-Alarm App}
\label{sec:smoke}
We present the Groovy source code of the \texttt{Smoke-Alarm} app in Listing~\ref{listing-Smoke-Alarm}, and its IR in Listing~\ref{listing-Smoke-Alarm-IR}. 

\begin{lstlisting}[caption=\texttt{Source code of Smoke-Alarm app}, label=listing-Smoke-Alarm]
/**
 *  Smoke-Alarm app
 *
 *  Author:Soteria
 */

definition(
    name: "SmartApp",
    namespace: "mygithubusername",
    author: "Model Analyzer",
    description: "Smoke-Detector App introduced in Section 3.",
    category: "Safety & Security",
    iconUrl: "https://s3.amazonaws.com/smartapp-icons/Convenience/Cat-Convenience.png",
    iconX2Url: "https://s3.amazonaws.com/smartapp-icons/Convenience/Cat-Convenience@2x.png",
    iconX3Url: "https://s3.amazonaws.com/smartapp-icons/Convenience/Cat-Convenience@2x.png")

preferences {
    section("Select smoke detector: "){
        input "smoke_detector", "capability.smokeDetector", title: "Which detector?", required: true
    }
    section("Select switch for low batter notification: "){
        input "the_switch", "capability.switch", title: "Which switch?", required: true
    }
    section("Select alarm device: ") {
        input "the_alarm", "capability.alarm", title: "Which alarm?", required: true
    }
    section("Select water valve: "){
        input "the_valve", "capability.valve", title: "Which valve?", required: true
    }
    section("Select battery settings: "){
        input "the_battery", "capability.battery", title: "Which battery?", required: true
    }
    section( "Low battery warning:  "){
        input "thrshld", "number", title: "Low Battery Threshold", required: true
    }
}

def installed()
{
	initialize()
}

def updated()
{
	unsubscribe()
	initialize()
}

private initialize() {
    subscribe(smoke_detector, "smoke", smokeHandler)
    subscribe(the_battery, "battery", batteryHandler)
}

def smokeHandler(evt) {
    log.trace "$evt.value: $evt, $settings"
    String theMessage
    log.debug "event created at: ${evt.date}"

    if (evt.value == "tested") {
    	theMessage = "${evt.displayName} tested for smoke."
    } else if (evt.value == "clear") {
    	theMessage = "${evt.displayName} is clear for smoke."
        the_alarm.off()
        the_valve.close()
        log.debug "evt clear"
    } else if (evt.value == "detected") {
        theMessage = "${evt.displayName} detected smoke!"
        the_alarm.siren()
        the_valve.open()
    } else {
    	theMessage = ("Unknown event received ${evt.name}")
    }
    log.warn "$theMessage"
}

def batteryHandler(evt) {
    log.trace "$evt.value: $evt, $settings"
    def String theMessage
    def check = thrshld
    def battLevel = findBatteryLevel()
   
    if (battLevel < check) {
        the_switch.on()
    	theMessage = "${evt.displayName} has battery ${battLevel}"
        log.warn "$theMessage -- ${batteryDevice.name} ${batteryDevice.label} battery: ${batteryLevel}% versus alarm at ${settings.threshold}"
    }
}

def findBatteryLevel(){
    log.trace "$evt.value: $evt, $settings"
    return the_battery.currentValue("battery").integerValue
}
\end{lstlisting}

\begin{lstlisting}[caption=\texttt{IR of Smoke-Alarm app}, label=listing-Smoke-Alarm-IR, keywordstyle=\color{black}]

/**
*  Smoke-Alarm app
*
*  Author:Soteria
*/
// IR of the Smoke-Alarm app

// Permissions block
input (smoke_detector, smokeDetector, type:device)
input (the_switch, switch, type:device)
input (the_alarm, alarm, type:device) 
input (the_valve, valve, type:device)
input (the_battery, battery, type:device)
input (thrshld, number, type:user_defined)

// Events/Actions block
subscribe(smoke_detector, "smoke", h1)
subscribe(the_battery, "battery", h2)

// Entry point
h1(){
  if(evt.value == "detected") {
      the_alarm.siren()
      the_valve.open()  
  } 
  if(evt.value=="clear"){
      the_alarm.off()
      the_valve.close()
  } 
}

// Entry point
h2(){
  check = thrshld
  batteryLevel = p()

  if(batteryLevel < check){
      the_switch.on()
   }
}

p(){
   return the_battery.currentValue("battery").integerValue
}

\end{lstlisting}

\subsection{Water Leak Detector App}
\label{sec:water}
We present the Groovy source code of the \texttt{Water\-Leak-Detector} app in Listing~\ref{listing-Water-Leak-Detector}, and its IR in Listing~\ref{listing-Water-Leak-Detector-IR}.

\begin{lstlisting}[caption=\texttt{Source code of Water-Leak-Detector app}, label=listing-Water-Leak-Detector]
/**
 *  Water-Leak-Detector app
 *
 *  Author:Soteria
 */
definition(
    name: "SmartApp",
    namespace: "mygithubusername",
    author: "Model Analyzer",
    description: "Water-Leak-Detector app introduced in Section 3. Additionally, it implements a push notification or text message for leak notification.",
    category: "Safety & Security",
    iconUrl: "https://s3.amazonaws.com/smartapp-icons/Meta/water_moisture.png",
    iconX2Url: "https://s3.amazonaws.com/smartapp-icons/Meta/water_moisture@2x.png"
)

preferences {
    section("When there's water detected...") {
        input "water_sensor", "capability.waterSensor", title: "Where?"
        input "valve_device", "capability.valve", title: "Valve device"
    }
    section("Send a notification to...") {
        input("recipients", "contact", title: "Recipients", description: "Send notifications to") {
           input "phone", "phone", title: "Phone number?", required: false
        }
    }
}

def installed(){
    subscribe(water_sensor, "water.wet", waterWetHandler)
}

def updated(){
    unsubscribe()
    subscribe(water_sensor, "water.wet", waterWetHandler)
}

def waterWetHandler(evt){
    def deltaSeconds = 60

    def timeAgo = new Date(now() - (1000 * deltaSeconds))
    def recentEvents = water_sensor.eventsSince(timeAgo)
    log.debug "Found ${recentEvents?.size() ?: 0} events in the last $deltaSeconds seconds"
    valve_device.close()
    def alreadySentSms = recentEvents.count {it.value && it.value == "wet"} > 1
    if (alreadySentSms){
        log.debug "SMS already sent within the last $deltaSeconds seconds"
	}else{
	    def msg = "${wsensor.displayName} is wet!"
	    log.debug "$wsensor is wet, texting phone number and closing valve"
	    if (location.contactBookEnabled){
	        sendNotificationToContacts(msg, recipients)
	    }
	    else{
	        sendPush(msg)
	        if (phone) {
	            sendSms(phone, msg)
	        }
	     }
	 }
}
\end{lstlisting}

\begin{lstlisting}[caption=\texttt{IR of Water-Leak-Detector}, label=listing-Water-Leak-Detector-IR, keywordstyle=\color{black}]
/**
*  Water-Leak-Detector app
*
*  Author:Soteria
*/
// IR of the Water-Leak-Detector app

// Permissions block
input (water_sensor, waterSensor, type:device)
input (valve_device, valve, type:device) 

//Events/Actions block
subscribe(water_sensor, "water.wet", h)

//Entry point
def h(evt){
    valve_device.close()
}

\end{lstlisting}

\subsection{Thermostat Energy Control App}
\label{sec:thermostat}

We present the Groovy source code of the \texttt{Thermostat\-Energy-Control} app in Listing~\ref{listing-Thermostat-Energy-Control}, and its IR in Listing~\ref{listing-Thermostat-Energy-Control-IR}.

\begin{lstlisting}[caption=\texttt{Source code of Thermostat-Energy-Control}, label=listing-Thermostat-Energy-Control]
/**
 *  Thermostat-Energy-Control app
 *
 *  Author:Soteria
 */

definition(
    name: "SmartApp",
    namespace: "mygithubusername",
    author: "Model Analyzer",
    description: "Smoke-Detector App introduced in Section 3. Additionally, the app warns the energy usage by text message",
    category:  "Green Living",
    iconUrl: "https://s3.amazonaws.com/smartapp-icons/Convenience/Cat-Convenience.png",
    iconX2Url: "https://s3.amazonaws.com/smartapp-icons/Convenience/Cat-Convenience@2x.png",
    iconX3Url: "https://s3.amazonaws.com/smartapp-icons/Convenience/Cat-Convenience@2x.png")

preferences {
    section("Control") {
        input "ther", "capability.thermostat", title: "Thermostat", required:true
    }
    section("Select the door lock:") {
        input "the_lock", "capability.lock", required: true
    }
    section("Select the thermostat energy meter to monitor:") {
        input "power_meter", "capability.powerMeter", title: "Energy Meters", required: true
        input "price_kwh", "number", title: "thereshold value for energy usage", required: true
    }
    section("Select the heater outlet switch:"){
        input "the_switch", "capability.switch", title: "Outlets", required: true
    }
   
    section( "Notifications" ) {
        input("recipients", "contact", title: "Send notifications to", required: false) {
        input "phoneNumber", "phone", title: "Warn with text message (optional)", description: "Phone Number", required: false
        }
    }

def installed(){
    log.debug "${app.label} installed with: ${settings}"
    initialize()
}

def updated(){
    unsubscribe()
    unschedule()
    initialize()
}

def initialize(){
    log.debug "Settings: ${settings}"
    subscribe(location, "mode", modeChangeHandler)
    subscribe(power_meter, "power", powerHandler)
}

def modeChangeHandler(evt) {
    log.trace "modeChangeHandler($evt.name: $evt.value)"
    def temp = 68
    setTemp(temp)
    the_lock.lock()
}


def setTemp(t){
    ther.setHeatingSetpoint(t)
    def msg = "heating and cooling point set, door is locked!"
    send(msg)
}

def powerHandler(evt){
    log.trace "$evt.displayName($evt.name:$evt.unit) $evt.value"
	
    def above_thrshld_val = 50 
    def below_thrshld_val = 5
    def dUnit = evt.unit ?: "Watts"
    
    power_val = get_power()
    
    if (power_val > above_thrshld_val ){
        def msg = "${power_meter} reported ${evt.value} ${dUnit} which is above the threshold of ${above_thrshld_val}."
        the_switch.off()
        send(msg)
    }
    if (power_val < below_thrshld_val ){
    	def msg = "${power_meter} reported ${evt.value} ${dUnit} which is below the threshold of ${below_thrshld_val}."
    	the_switch.on()
    	send(msg)
    } 
}

def get_power(){
    log.debug energy_meter.currentValue('energy')
    latest_power = power_meter.currentValue("power")
    return latest_power	
}

def send(msg){
    if(location.contactBookEnabled) {
        if (recipients) {
            log.debug ( "Sending Push Notification..." ) 
            sendNotificationToContacts(msg, recipients)
        }
    }
    if (phoneNumber) {
        log.debug("Sending text message...")
        sendSms( phoneNumber, msg)
    }
}
\end{lstlisting}

\begin{lstlisting}[caption=\texttt{IR of Thermostat-Energy-Control}, label=listing-Thermostat-Energy-Control-IR, keywordstyle=\color{black}]
/**
*  Thermostat-Energy-Control app
*
*  Author:Soteria
*/
// IR of the Thermostat-Energy-Control app

// Permission block 
Input(ther, thermostat, type:device )
Input(power_meter, powerMeter, type:device)
Input(the_switch, switch, type:device)
Input(the_lock, lock, type:device)

// Event/Action block
subscribe(location, mode, h1)
subscribe(power_meter, power, h2)

// Entry point
h1(){
   def temp = 68
   the_lock.lock()
   setTemp(temp)
}

// Entry point 
h2(){
   above_thrshld_val = 50 
   below_thrshld_val = 5

   power_val = get_power()

   if (power_val > above_thrshld_val ){
      the_switch.off()
   }
   if (power_val < below_thrshld_val ){
      the_switch.on()
   }
}

def setTemp(t){
   ther.setHeatingSetpoint(t)
}

get_power(){
   latest_power = power_meter.currentValue("power")
   return latest_power	
}
\end{lstlisting}

\section{Properties of IoT Apps}
\label{appendix:properties}
We present general IoT properties in Table~\ref{table:general_properties} and application-specific IoT properties in Table~\ref{table:application_properties}. 

 \makeatletter
    \setlength\@fptop{0\p@}
\makeatother 

\begin{table}[t!]
\centering
\newcolumntype{P}[1]{>{\centering\arraybackslash}p{#1}}
\small
\def\arraystretch{1.1}
\begin{tabular}{|P{0.4cm}|p{6.3cm}|}
\hline
\multicolumn{1}{|c|}{\textbf{ID}} & \multicolumn{1}{c|}{\textbf{ Description\tnote{\textdaggerdbl}}} \\ \hline \hline
$\mathtt{\textbf{S.1}}$ & An event handler must not change a device attribute to conflicting values on some control-flow path, \eg the motion-active event handler must not turn on and turn off a switch in some branch. \\ \hline
$\mathtt{\textbf{S.2}}$ & An event handler must not change a device attribute to the same value multiple times on some control-flow path, \eg the motion-active event handler must not turn on the switch multiple times in some branch. \\ \hline
$\mathtt{\textbf{S.3}}$ & Event handlers of complement events must not change a device attribute to the same value, \eg the motion-active event handler and the motion-inactive event handler must not both turn on a switch. \\ \hline
$\mathtt{\textbf{S.4}}$ & Two or more non-complement event handlers must not change a device attribute to conflicting values, \eg a user-present event handler turns on the switch while a timer event handler turns off the switch at midnight. This is because the events of user presence and midnight may occur at the same time, leading to a race condition. \\ \hline
$\mathtt{\textbf{S.5}}$ & An event must be subscribed by the event handler whose code contains logic that handles that event. A violation happens when (1) a handler takes an event-typed value and performs different actions according to the types of events, and (2) the handler has a case for handling event $e$, but (3) the app does not declare that the handler subscribes to event $e$. For example, a handler checks for the motion-active event and turns on a switch when the motion is active, but the app does not declare that the handler subscribes to the motion-active event.
\\ \hline
\end{tabular}%
\caption{Description of general properties.}
\label{table:general_properties}
\end{table}

\begin{table*}[ht!]
\newcolumntype{P}[1]{>{\centering\arraybackslash}p{#1}}
\def\arraystretch{1.4}
\centering
\footnotesize
\begin{threeparttable}[b]
\begin{tabular}{|P{0.5cm}|p{14.5cm}|}
\hline
\multicolumn{1}{|c|}{\textbf{ID\textdagger}} & \multicolumn{1}{c|}{\textbf{Property Description\tnote{\textdaggerdbl}}} \\ \hline \hline
$\mathtt{\textbf{P.1}}$& The door must be locked when a user is not present at home or sleeping. \\ \hline
$\mathtt{\textbf{P.2}}$ & The lights (in a bedroom, hallway, etc.) must be turned on if the motion sensor is active. \\ \hline
$\mathtt{\textbf{P.3}}$ & When there is smoke, the lights must be on if it is night, and the door must be unlocked. \\ \hline
$\mathtt{\textbf{P.4}}$& The light must be on when the user arrives home.  \\ \hline
$\mathtt{\textbf{P.5}}$& The camera controlled doors must be closed when the door is clear of any objects. \\ \hline
$\mathtt{\textbf{P.6}}$ & The garage door must be open when people arrive home, and it must be closed when people leave home. \\ \hline
$\mathtt{\textbf{P.7}}$ & The location beacon must be inside a geofence around the home (defined by a user) to turn on the lights and open the garage door. \\ \hline
$\mathtt{\textbf{P.8}}$ & The lights must be turned off when the sleep sensor detects a user is sleeping. \\ \hline
$\mathtt{\textbf{P.9}}$ & The security system must not be disarmed when the user is not at home. \\ \hline
$\mathtt{\textbf{P.10}}$ & The alarm must sound when there is smoke or CO; and when an unexpected motion, tampering, and entering occurs. \\ \hline
$\mathtt{\textbf{P.11}}$ & The valve must be closed when water sensor is wet and when the water level threshold specified by a user is reached. \\ \hline
$\mathtt{\textbf{P.12}}$ & The devices (\eg light switches, cleaning supply cabinets, medicine drawers, or gun cases) must not be open or turned on when the user is not at home or sleeping. \\ \hline
$\mathtt{\textbf{P.13}}$ & Some device functionality (\eg coffee machine starting brewing, heating up dinner in a crock-pot) must not be used when the user is not at home or must be turned on before a time specified by a user. \\ \hline
$\mathtt{\textbf{P.14}}$ & The refrigerator, alarm, and security system must not be disabled, and their use must not be restricted to save energy. \\ \hline
$\mathtt{\textbf{P.15}}$& The temperature must be set to the operating mode values as defined by the user (heat mode and cool mode are separate) when there is motion; when there is no motion the idle temp (energy savings) must be set as defined by the user (heat mode and cool mode are separate). \\ \hline
$\mathtt{\textbf{P.16}}$ & The thermostat temperature (heating and cooling) entered by the user must be changed when the mode selected by a user is changed (\eg from sleeping mode to away mode). \\ \hline
$\mathtt{\textbf{P.17}}$& The AC and heater must not be on at the same time. \\ \hline
$\mathtt{\textbf{P.18}}$ & The HVACs, fans, switches, heaters, dehumidifiers must be off when the temperature and humidity are out of the zone entered by the user (\eg one degree above/below the temperature threshold and 5\% below the humidity threshold). \\ \hline
$\mathtt{\textbf{P.19}}$ & The AC must be on when a user is within a specified distance of the house or at a time specified by the user. \\ \hline
$\mathtt{\textbf{P.20}}$ & The security camera must take pictures when there is a motion, and contact/door sensors are active. \\ \hline
$\mathtt{\textbf{P.21}}$& The security camera must take a photo and sound alarm when doors/windows are opening, and doors are unlocking during user-specified times. It must turn off all alarm when one alarm is turned off. \\ \hline
$\mathtt{\textbf{P.22}}$ & The battery level of the devices (switch, humidity sensor, etc.) must not be below a specified threshold. \\ \hline
$\mathtt{\textbf{P.23}}$ & The door must not be unlocked when a camera does not recognize an unauthorized face. \\ \hline
$\mathtt{\textbf{P.24}}$ & The windows must not be open when the heater is on. \\ \hline
$\mathtt{\textbf{P.25}}$ & The bell must not chime when the door is closed. \\ \hline
$\mathtt{\textbf{P.26}}$ & The alarm must go off when the main door is left open for too long (specified by the user). \\ \hline
$\mathtt{\textbf{P.27}}$ & The mode must be set to ``home'' when the user is at home, and ``away'' when the user is not at home. \\ \hline
$\mathtt{\textbf{P.28}}$ & The sound system must read (\eg the day's weather forecast and the status of the devices) with the user interaction and at the time not specified by the user (guards against violations when the sleeping mode is on and when the user is not home.)  \\ \hline
$\mathtt{\textbf{P.29}}$ & The sprinkler system must not be on when it is raining, and the soil moisture is below a threshold defined by a user. Flood sensor must activate the alarm when there is water. \\ \hline
$\mathtt{\textbf{P.30}}$ & The water valve must shut off when water/moisture sensor detects a leak around a location such as a basement and a laundry room. \\ \hline
\end{tabular}
\begin{tablenotes}
    \item [\textdagger] \footnotesize{We define general properties based on the access granted to the devices in an app. For instance, $\mathtt{P.22}$ is separately defined for an app that grants access to a switch and a humidity sensor.}
    \vspace{2pt}
    \item[\textdaggerdbl] \footnotesize{The properties are validated in some cases with the predicates provided by users or developers. For instance, for $\mathtt{P.4}$, the smart porch light must turn on when a user enters a geofence of a house at a specific time defined by a user;  for $\mathtt{P.21}$, the battery level of devices is checked at a time specified by a user or a default time defined by the developer.}
\end{tablenotes}
\end{threeparttable}
\caption{Description of application-specific properties.}
\label{table:application_properties}
\end{table*}

\section{\textsc{MalIoT} Test Suite Applications}
\label{appendix:iotmalicious_apps}
Table~\ref{appendix:iotmalicious-table} presents \MalIoT apps categorized by their property violation ground-truth. The source code of the \MalIoT apps are available at IoTBench repository~\cite{iotbench}

\begin{table*}[ht!]
\newcolumntype{P}[1]{>{\centering\arraybackslash}p{#1}}
\def\arraystretch{1.2}
\centering
\footnotesize
\begin{threeparttable}[b]
\begin{tabular}{|P{0.8cm}|p{5cm}|p{5cm}|P{2cm}|P{0.8cm}|}
\hline
\multicolumn{1}{|c|}{\textbf{ID}} & \multicolumn{1}{c|}{\textbf{Description}} & \multicolumn{1}{c|}{\textbf{Pr. Violation}} & \multicolumn{1}{c|}{\textbf{Imp. Details\tnote{\textdaggerdbl}}} & \multicolumn{1}{c|}{\textbf{Results}} \\ \hline \hline
$\mathtt{\textbf{App1}}$ & The lights are turned off at night when motion is detected. & $\mathtt{P.2}$ is violated. The app prevents brightening the path the user is walking. & Device events &  \CheckmarkBold  P\\ \hline
$\mathtt{\textbf{App2}}$& The security system is turned off when there is nobody at home. & $\mathtt{P.9}$ is violated. The app could leave the house vulnerable to break-ins. & State variables, predicate analysis & \CheckmarkBold P \\ \hline
$\mathtt{\textbf{App3}}$ & A battery operated switch is turned off every 30 seconds. & $\mathtt{S.2}$ is violated. This is similar to DDoS attacks to consume the battery of the switch by sending the same command to the device multiple times. & Device events, timer events & \CheckmarkBold S \\ \hline
$\mathtt{\textbf{App4}}$ & The app turns off a switch of a device to save energy after a number of minutes specified by user. However, the app keeps the device turned on. & $\mathtt{S.1}$ is violated. The event handler changes conflicting attributes of a switch device (switch on and switch off). & Device events, multiple entry points & \CheckmarkBold S \\ \hline
$\mathtt{\textbf{App5}}$ & The app sounds the alarm when there is smoke.It also implements another method that turns off the alarm when there is smoke. & A string is used to invoke a method via call by reflection. A method violates $\mathtt{P.2}$ which turns off the alarm when there is smoke. & Call by reflection, state variables &  \textbf{\text{\sffamily X}}  \\ \hline
$\mathtt{\textbf{App6}}$ & When a user leaves home, the light illuminance level changed from 0 to a numerical value, and the door is unlocked after some time. & $\mathtt{P.1}$ and $\mathtt{P.13}$ are violated. This allows an attacker to be aware that the user is not at home, and could let the attacker break into the home. & Multiple violations, multiple entry points, timer events & \CheckmarkBold P \\ \hline
$\mathtt{\textbf{App7}}$& The app turns on and turns off switches at a specified time defined by a user. Additionally, it turns on the switches when the user is at home and turns them off when the user is not at home. & $\mathtt{S.4}$ is violated. The user's presence to turn on the switch and the time entered by the user to turn off the switch may happen at the same time. & Multiple Entry points, timer events & \CheckmarkBold S \\ \hline
$\mathtt{\textbf{App8}}$ & The app does not subscribe the location mode change event handler to lock the door when the user is away or unlock the door when the user is at home. & $\mathtt{S.5}$ and $\mathtt{P.1}$ is violated. The app fails to invoke the location mode change event handler method which fails to lock or unlock the door. & Multiple violations, multiple entry points, predicate analysis, mode events & \CheckmarkBold P S\\ \hline
$\mathtt{\textbf{App9}}$ & Location mode is set to home when the user is not at home. & $\mathtt{P.27}$ is violated. A string is requested via \texttt{HttpGet} interface and used in a call by reflection.  A run-time analysis is required to check whether the method is invoked via the string. & Call by reflection & \textbf{\text{\sffamily O}} \\ \hline
$\mathtt{\textbf{App10}}$ & The app uses dynamic permissions based on previous selections or external inputs to control a set of devices. & The app dynamically generates the content of a page where the device the permissions of the devices  depends on the previously selected device permissions. & Dynamic device permissions & \textbf{\text{\sffamily !}} \\ \hline
$\mathtt{\textbf{App11}}$ & The app sends a notification to the user when the kids leave home. & The app also notifies the attacker via \texttt{sendSms} interface. & Multiple sensitive data leaks & \textbf{\text{\sffamily !}} \\ \hline
$\mathtt{\textbf{App12}}$ & The app turns on the light switches when the alarm sounds. & \multirow{3}{5cm}{\\$\mathtt{P.3}$ is violated.  $\mathtt{App12}$,  $\mathtt{App13}$, and  $\mathtt{App14}$ interact each other, and locks the door when there is smoke in the house (we note that the use of an individual app does not violate any properties.)} & \multirow{3}{2cm}{\\\centering Predicate analysis, device events, mode events} & \multirow{3}{0.5cm}{\\\centering \CheckmarkBold P} \\ \cline{1-2}
$\mathtt{\textbf{App13}}$ & The app changes the mode from away to home when the light switch is turned on so that it is aware that user is at home. &  &  &  \\ \cline{1-2}
$\mathtt{\textbf{App14}}$ & The app locks the door when the home mode is triggered. &  &  &  \\ \hline
$\mathtt{\textbf{App15}}$ & The lights are turned off when motion is detected. & $\mathtt{S.1}$ is violated. This app interacts with $\mathtt{App1}$ and invokes the conflicting attributes of the same device (lights on and lights off). & Device events & \CheckmarkBold S\\ \hline
$\mathtt{\textbf{App16}}$ & The app changes mode to sleeping when the user turns off the bedroom lights. & \multirow{2}{5cm}{$\mathtt{P.14}$ is violated multiple times. $\mathtt{App16}$ and  $\mathtt{App17}$ interact with each other. This allows the sleeping mode change event to turn off the alarm and security camera.} & \multirow{2}{2cm}{\centering Device events, mode events} & \multirow{2}{0.5cm}{\centering \CheckmarkBold P} \\ \cline{1-2}
$\mathtt{\textbf{App17}}$ & The app turns off all plugged devices when the sleeping mode is triggered. &  &  &  \\ \hline
\end{tabular}
\begin{tablenotes}
    \item [\textdagger] The details present the Groovy language- and IoT- specific properties that require program analysis for the verification of properties.
        \vspace{2pt}
    \item[\textdaggerdbl]  \CheckmarkBold P = True Positive (model checked with $\mathtt{P1}$-$\mathtt{P30}$),  \CheckmarkBold S= True Positive (model checked with $\mathtt{S1}$-$\mathtt{S5}$), \textbf{\text{\sffamily X}} = False Positive, \textbf{\text{\sffamily O}} = Dynamic analysis required,  \textbf{\text{\sffamily !}} = Not considered in attacker model
\end{tablenotes}
\end{threeparttable}
\caption{Description of \MalIoT test suite apps and \Soteria's results.}
\label{appendix:iotmalicious-table}
\end{table*}

}

\end{document}